\documentclass[11pt,a4paper,twoside,groupcitations]{article}
\usepackage[T1]{fontenc}
\usepackage[ansinew]{inputenc}
\usepackage[english]{babel}
\usepackage{amsfonts}
\usepackage{amsmath}
\usepackage{array}
\usepackage{amsthm}
\usepackage{amssymb}
\usepackage{graphicx}
\usepackage{subfigure}
\usepackage{braket}
\usepackage{eucal}
\usepackage{verbatim}
\usepackage[table]{xcolor}
\usepackage{caption}
\usepackage{cite}
\usepackage{textcomp}
\usepackage{multicol}
\usepackage{tikz}
\usetikzlibrary{positioning,arrows}
\usetikzlibrary{decorations.pathmorphing}
\usetikzlibrary{decorations.markings}
\usetikzlibrary{calc,decorations.markings}
\usetikzlibrary{arrows,shapes}
\usetikzlibrary{matrix,arrows}
\usepackage{pgfplots}
\usepackage{xparse}
\definecolor{jade}{HTML}{00A86B}
\newcommand{\be}{\begin{eqnarray}}
\newcommand{\ee}{\end{eqnarray}}

\renewcommand{\d}{\mbox{${\rm d}$}} 

\newcommand{\gn}{G_{\rm N}}
\newcommand{\rh}{r_{\rm H}}

\title{\bf Polytropic stars in bootstrapped Newtonian gravity}
\author{Roberto~Casadio$^{ab}$\thanks{E-mail: casadio@bo.infn.it}
$\ $
and
Octavian Micu$^c$\thanks{E-mail: octavian.micu@spacescience.ro}
\\
\\
$^a${\em Dipartimento di Fisica e Astronomia, Universit\`a di Bologna}
\\
{\em via Irnerio~46, 40126 Bologna, Italy}
\\
\\
$^b${\em I.N.F.N., Sezione di Bologna, I.S.~FLAG}
\\
{\em viale B.~Pichat~6/2, 40127 Bologna, Italy}
\\
\\
{\em $^b$Institute of Space Science, Bucharest, Romania}
\\
{\em P.O. Box MG-23, RO-077125 Bucharest-Magurele, Romania}
}
\begin{document}
\maketitle
\begin{abstract}
We study self-gravitating stars in the bootstrapped Newtonian picture for polytropic equations of state. 
We consider stars that span a wide range of compactness values.
Both matter density and pressure are sources of the gravitational potential. 
Numerical solutions show that the density profiles can be well approximated by Gaussian functions.
Later we assume Gaussian density profiles to investigate the interplay between the compactness
of the source, the width of the Gaussian density profile and the polytropic index. 
We also dedicate a section to comparing the pressure and density profiles of the bootstrapped Newtonian
stars to the corresponding General Relativistic solutions.
We also point out that no Buchdahl limit is found, which means that the pressure can in principle
support a star of arbitrarily large compactness.
In fact, we find solutions representing polytropic stars with compactness above
the Buchdhal limit.
%
\end{abstract}
\section{Introduction and motivation}
\setcounter{equation}{0}
\label{Sintro}
One of the most striking features of the strong gravity regime in General Relativity 
is that, once a trapping surface appears, singularity theorems require an object to collapse
all the way into a region of infinite density surrounded by a black hole 
geometry~\cite{HE}.
Static black hole spacetimes are however problematic in this classical description,
since point-like sources are mathematically incompatible with the Einstein equations~\cite{geroch}.
One would therefore hope that quantum physics solves this fundamental puzzle in the description of
self-gravitating objects, the same way it removes the ultraviolet catastrophe and makes the hydrogen atom stable. 
Given the strong experimental constraints on possible deviations from General Relativity, quantum
effects can only become significant in the strong field regime, where perturbative methods hardly
apply and matter likely requires physics beyond the standard model as well~\cite{brustein}.
In particular, whether the scale at which quantum departures from General Relativity become appreciable
is significantly large to affect the description of compact astrophysical objects remain a key physical question. 
\par
In light of the above observations, in Refs.~\cite{Casadio:2019cux,BootN} we studied solutions of an effective
equation for the gravitational potential of a static source which contains a gravitational self-interaction
term besides the usual Newtonian coupling with the matter density.
Following an idea from Ref.~\cite{Casadio:2016zpl}, the self-interaction term was derived in details
from a Fierz-Pauli Lagrangian in Ref.~\cite{Casadio:2017cdv}, and it could therefore be viewed
as the first step in the perturbative reconstruction of General Relativity (see {\em e.g.}, Refs.~\cite{carballo}).
However, since the equation for the potential was solved non-perturbatively~\cite{Casadio:2019cux,BootN},
it could also be conjectured that this ``bootstrapped'' Newtonian gravity effectively describes the (mean field)
quantum gravitational potential of extremely compact objects after the break-down of classical
General Relativity~\cite{Casadio:2020mch,kuntz}.
Moreover, we found no equivalent of the Buchdahl limit~\cite{buchdahl}, a result implying that matter pressure
(possibly of quantum origin) could support sources of arbitrarily large compactness
\be
X
\equiv
\frac{\gn\,M}{R}
\ ,
\ee
where $R$ is the radius and $M$ the ADM-like~\cite{adm} mass~\cite{Casadio:2019pli} of the source.~\footnote{In this
paper we use units with $c=1$, and always display the Newton constant $\gn$ explicitly.} 
\par
Like in Refs.~\cite{Casadio:2019cux,BootN}, we shall just consider (static) spherically symmetric systems,
so that all quantities depend only on the radial coordinate $r$, but the density profile is not restricted to
be uniform.
We shall begin by assuming that the matter density $\rho=\rho(r)$ and pressure $p=p(r)$ satisfy a polytropic
equation of state and determine the density profile numerically for different values of the compactness
(and of the polytropic parameters).
This analysis will show that the equilibrium configurations closely resemble Gaussian distributions.
Therefore, we shall also study Gaussian density profiles analytically and determine {\em a posteriori\/} the 
compatible polytropic parameters. 
\par
The paper is organised as follows:
in Section~\ref{S:action}, we briefly review the bootstrapped Newtonian picture;
Section~\ref{S:solution} is dedicated to the investigation of the model within the further assumption of a polytropic
equation of state for the matter source and to finding numerical solutions for the density profile;
that is followed in Section~\ref{S:gaussian} by an in depth analysis of Gaussian density profiles;
finally we comment about our results and possible outlooks in Section~\ref{S:conc}.
\section{Bootstrapped theory for the gravitational potential}
\label{S:action}
\setcounter{equation}{0}
From Ref.~\cite{Casadio:2017cdv}, we recall that the non-linear equation for the potential $V=V(r)$
describing the gravitational pull on test particles generated by a matter density $\rho=\rho(r)$ can be
obtained starting from the Newtonian Lagrangian
\be
L_{\rm N}[V]
=
-4\,\pi
\int_0^\infty
r^2 \,\d r
\left[
\frac{\left(V'\right)^2}{8\,\pi\,\gn}
+\rho\,V
\right]
\ ,
\label{LagrNewt}
\ee
where $f'\equiv\d f/\d r$, and the corresponding field equation is the Poisson equation
\be
r^{-2}\left(r^2\,V'\right)'
\equiv
\triangle V
=
4\,\pi\,\gn\,\rho
\label{EOMn}
\ee
for the Newtonian potential $V=V_{\rm N}$.
We can then include the effects of gravitational self-interaction by noting that the Hamiltonian
\be
H_{\rm N}[V]
=
-L_{\rm N}[V]
=
4\,\pi
\int_0^\infty
r^2\,\d r
\left(
-\frac{V\,\triangle V}{8\,\pi\,\gn}
+\rho\,V
\right)
\ ,
\label{NewtHam}
\ee 
computed on-shell by means of Eq.~\eqref{EOMn}, yields the Newtonian potential energy
\be
U_{\rm N}(r)
&\!\!=\!\!&
2\,\pi
\int_0^r 
{\bar r}^2\,\d {\bar r}
\,\rho(\bar r)\, V(\bar r)
\nonumber
\\
&\!\!=\!\!&
-\frac{1}{2\,\gn}\,
\int_0^r 
{\bar r}^2 \,\d {\bar r}\,
\left[V'(\bar r) \right]^2
\ ,
\label{Unn}
\ee
where we used Eq.~\eqref{EOMn} and then integrated by parts discarding boundary terms.
One can view the above $U_{\rm N}$ as given by the interaction of the matter distribution enclosed in a sphere
of radius $r$ with the gravitational field.
Following Ref.~\cite{Casadio:2016zpl} (see also Refs.~\cite{dadhich1}), we then define a self-gravitational
source proportional to the gravitational energy $U_{\rm N}$ per unit volume, that is 
\be
J_V
\simeq
\frac{\d U_{\rm N}}{\d \mathcal{V}} 
=
-\frac{\left[ V'(r) \right]^2}{2\,\pi\,\gn}
\ .
\label{JV}
\ee
In Ref.~\cite{BootN}, we found that the pressure $p$ becomes very large for compact sources
with $X\gtrsim 1$, and we must therefore add a corresponding potential energy $U_{\rm B}$ such that
\be
p
=
-\frac{\d U_{\rm B}}{\d \mathcal{V}} 
\ .
\label{JP}
\ee
Since the latter contribution just adds to $\rho$, it can be easily included by simply shifting 
$\rho \to \rho+q_c\,p$, where $q_c$ is a positive constant which allows us to implement the
non-relativistic limit formally as $q_c\to 0$.
Upon including these new source terms, and the analogous higher order term $J_\rho=-2\,V^2$
which couples with the matter source, we obtain the total Lagrangian~\cite{Casadio:2017cdv}  
\be
L[V]
&\!\!=\!\!&
L_{\rm N}[V]
-4\,\pi
\int_0^\infty
r^2\,\d r
\left[
q_\Phi\,J_V\,V
+
q_\Phi\,J_\rho\left(\rho+q_c\,p\right)
\right]
\nonumber
\\
&\!\!=\!\!&
-4\,\pi
\int_0^\infty
r^2\,\d r
\left[
\frac{\left(V'\right)^2}{8\,\pi\,\gn}
\left(1-4\,q_\Phi\,V\right)
+\left(\rho+q_c\,p\right) V\left(1-2\,q_\Phi\,V\right)
\right]
\ ,
\label{LagrV}
\ee
where the positive parameter $q_\Phi$ plays the role of a coupling constant
for the graviton current $J_V$ and the higher-order matter current $J_\rho$.  
The associated effective hamiltonian is simply given by
\be
H[V]
=
-L[V]
\ .
\label{HamV}
\ee
Finally, the Euler-Lagrange equation for $V$ is given by
\be
\triangle V
=
4\,\pi\,\gn\left(\rho+q_c\,p\right)
+
\frac{2\,q_\Phi\left(V'\right)^2}
{1-4\,q_\Phi\,V}
\label{EOMV}
\ee
and the conservation equation that determines the pressure reads
\be
p'
=
-V'\left(\rho+q_c\,p\right)
\ .
\label{eqP}
\ee
The exact Newtonian equations are then recovered by taking the non-relativistic limit as $q_c\to 0$
and switching off the graviton self-interaction with $q_\Phi\to 0$.
It is important to remark that the couplings $q_\Phi$ and $q_c$ do not need to be small.
In fact, the closest results to General Relativity are expected to occur for
$q_\Phi\simeq q_c\simeq 1$~\cite{Casadio:2019cux}.
\section{Polytropic stars}
\label{S:solution}
\setcounter{equation}{0}
In Refs.~\cite{Casadio:2017cdv,BootN,Casadio:2019cux,Casadio:2019pli}, the effects of the gravitational self-interaction,
encoded by the term proportional to $q_\Phi$ in the field equation~\eqref{EOMV}, were analysed by considering simple sources,
characterised by a homogeneous matter density $\rho=\rho_0$ and different values of the compactness $X$.
One of the main results is that, for a flat density profile, the outer mass parameter $M$ is always larger than the proper mass
\be
M_0
=
m(R)
=
4\,\pi
\int_0^R
r^2\,\d r\,\rho(r)
\ .
\label{proper_mass}
\ee
This is in agreement with the fact that $M$ should also account for the (positive) pressure required to ensure equilibrium,
with $M$ approaching $M_0$ for smaller and smaller compactness $X$~\cite{Casadio:2019cux}.
\par
We now want to study more realistic matter distributions, for which we expect that the degeneracy pressure is the main
component counteracting the gravitational pull, like in neutron stars and white dwarfs.
For this purpose, we will assume a polytropic equation of state~\cite{polytropes} 
\be
p(r)
=
\gamma\, \rho^n(r)
=
\tilde\gamma\,\rho_0
\left[
\frac{\rho(r)}{\rho_0}
\right]^{n}
\ ,
\label{eosB}
\ee
with $n$ and $\tilde \gamma$ positive dimensionless parameters, and $\rho_0\equiv \rho(0)$ is used as a reference density.
Moreover, we shall also assume the surface pressure vanishes, $p_R\equiv p(R)=0$, which then implies that $\rho_R\equiv \rho(R)=0$.
\par
The relevant solutions for the density profile will have to lead to a potential which satisfies
the regularity condition in the centre
\be
V_{\rm in}'(0)=0
\label{b0}
\ee
and be smooth across the surface $r=R$, that is
\be
V_{\rm in}(R)
&\!\!=\!\!&
V_{\rm out}(R)
\equiv
V_R
\label{bR}
\\
\notag
\\
V'_{\rm in}(R)
&\!\!=\!\!&
V'_{\rm out}(R)
\equiv 
V'_R
\ ,
\label{dbR}
\ee
where we defined $V_{\rm in}=V(0\le r\le R)$ and $V_{\rm out}=V(R\le r)$.
The consistency of these two conditions with Eq.~\eqref{eosB} will be thoroughly analysed below after
we recall the outer vacuum solution.
\subsection{Outer vacuum solution}
\label{S:vacuum}
Outside the source, we have $\rho=p=0$ and Eq.~\eqref{eqP} is trivially satisfied.
Eq.~\eqref{EOMV} reads
\be
\triangle V
=
\frac{2\,q_\Phi \left(V'\right)^2}{1-4\,q_\Phi\,V}
\ ,
\label{EOMV0}
\ee
which is exactly solved by
\be
V_{\rm out}
=
\frac{1}{4\,q_\Phi}
\left[
1-\left(1+\frac{6\,q_\Phi\,\gn\,M}{r}\right)^{2/3}
\right]
\ .
\label{sol0}
\ee
where two integration constants were fixed by requiring the expected
Newtonian behaviour in terms of the ADM-like mass $M$ for large $r$. 
In fact, for large $r$, we have
\be
V_{\rm out}
\simeq
-\frac{\gn\,M}{r}
+q_\Phi\,\frac{\gn^2\,M^2}{r^2}
-q_\Phi^2\,\frac{8\,\gn^3\,M^3}{3\,r^3}
\ ,
\label{Vlarge}
\ee
which displays the expected post-Newtonian term of order $\gn^2$ for $q_\Phi=1$~\cite{Casadio:2017cdv}.
\par
For sufficiently large $X$, the outer potential gives rise to a ``Newtonian'' horizon of radius~\cite{Casadio:2019cux}
\be
\rh
\simeq
1.4\,\gn\,M
\ ,
\ee
precisely where $2\,V_{\rm out}(\rh)=-1$.
In this work, we shall therefore assume $X<0.7$ in order to avoid this feature.
\par
From Eq.~\eqref{sol0}, we also obtain 
\be
V_R
=
V_{\rm out}(R)
=
\frac{1}{4\,q_\Phi}
\left[
1-\left(1+6\,q_\Phi\,X\right)^{2/3}
\right]
\ ,
\label{VR} 
\ee
and
\be
R\,V'_R
=
R\,V_{\rm out}'(R)
=
\frac{X}
{\left(1+6\,q_\Phi\,X\right)^{1/3}}
\ ,
\label{dVR}
\ee
which we will often use since they appear in the boundary conditions~\eqref{bR} and \eqref{dbR}.
\subsection{The inner pressure and potential}
The conservation equation~\eqref{eqP} for the polytropic equation of state~\eqref{eosB} immediately allows one
to find the derivative of the potential
\be
V'
=
-\frac{n\,\gamma\,\rho'}{q_c\,\gamma\,\rho+\rho^{\,2-n}}
=
-\frac{n\,\tilde\gamma\,\rho'}{q_c\,\tilde\gamma\,\rho+\rho_0^{n-1}\,\rho^{2-n}}
\ .
\label{VR_rho1}
\ee
The regularity condition~\eqref{b0} then requires $\rho'(0)=0$, which can hold for any $\rho_0\equiv\rho(0)>0$. 
For $n\not= 1$, the above equation yields
\be
V
=
\beta
-\frac{n}{(n-1)\,q_c}\,\ln \left(1+q_c\,\gamma\,\rho^{n-1} \right)
=
\beta
-\frac{n}{(n-1)\,q_c}\,\ln \left[1+q_c\,\tilde\gamma\left(\frac{\rho}{\rho_0}\right)^{n-1} \right]
\ ,
\label{VR_rho}
\ee
where $\beta$ is an integration constant.
On the other hand, for $n=1$ the constant $\gamma=\tilde\gamma$ is dimensionless, and we find the simpler solution
\be
V
=
\beta
-\frac{\gamma}{q_c\,\gamma+1}\,\ln\left(\frac{\rho}{\rho_0}\right)
\ .
\label{Vn1}
\ee
We remark that Eqs.~\eqref{VR_rho} and~\eqref{Vn1} reproduce the Newtonian behaviour in the non-relativistic
limit $q_c\to 0$.
Moreover, the above expressions for $V$ and $V'$ evaluated at $r=R$ must equal the respective outer
values~\eqref{VR} and \eqref{dVR}, which only depend on $M$ and $R$, but not on any equation
of state.
Let us then analyse in details under which conditions the equation of state~\eqref{eosB} is compatible
with the continuity of the potential and its derivative.
\par
For $n>1$, continuity of the potential~\eqref{VR_rho} across $r=R$ simply fixes the integration constant 
\be
\beta
=
V_R
\ ,
\label{beta_0}
\ee
where we used $\rho_R=0$.
Values of $n$ outside this range must however be excluded.~\footnote{Eq.~\eqref{eosB} is often written
as $p=\gamma\,\rho^{1+1/n'}$, with $n'>0$ in the astrophysical literature.}
In fact, for $n=1$, the solution~\eqref{Vn1} diverges positively and so does Eq.~\eqref{VR_rho}
for $0<n\le 1$.
\par
Moreover, in the allowed range $n>1$, continuity of the derivative of the potential demand
\be
\frac{V_R'}{n\,\gamma}
=
-\lim_{r\to R}
\left(
\frac{\rho'}
{\rho^{2-n}}
\right)
\equiv
\frac{Y}{n\,\gamma\,R}
>0
\ ,
\ee
which implies that $\rho'\sim -\rho^{2-n}$ for $r\to R$.
It is important to remark that this condition holds irrespectively of the value of $q_c$, precisely because 
$n>1$ implies that the term $q_c\,\rho$ vanishes faster than $\rho^{2-n}$.
For $1<n<2$, both $\rho$ and $\rho'$ must then vanish at the star surface.
For $n=2$, the derivative $\rho'$ must be finite there, whereas for $n>2$ it must diverge (negatively)
for $r\to R$ from inside.
This latter behaviour could be roughly approximated with a step discontinuity at the surface of the star. 
We then remark that the text-book cases of relativistic ($n=4/3$) and non-relativistic ($n=5/3$) fermions
belong to the range $1<n<2$.
\par
\begin{figure}[t]
\centering
\includegraphics[width=10cm]{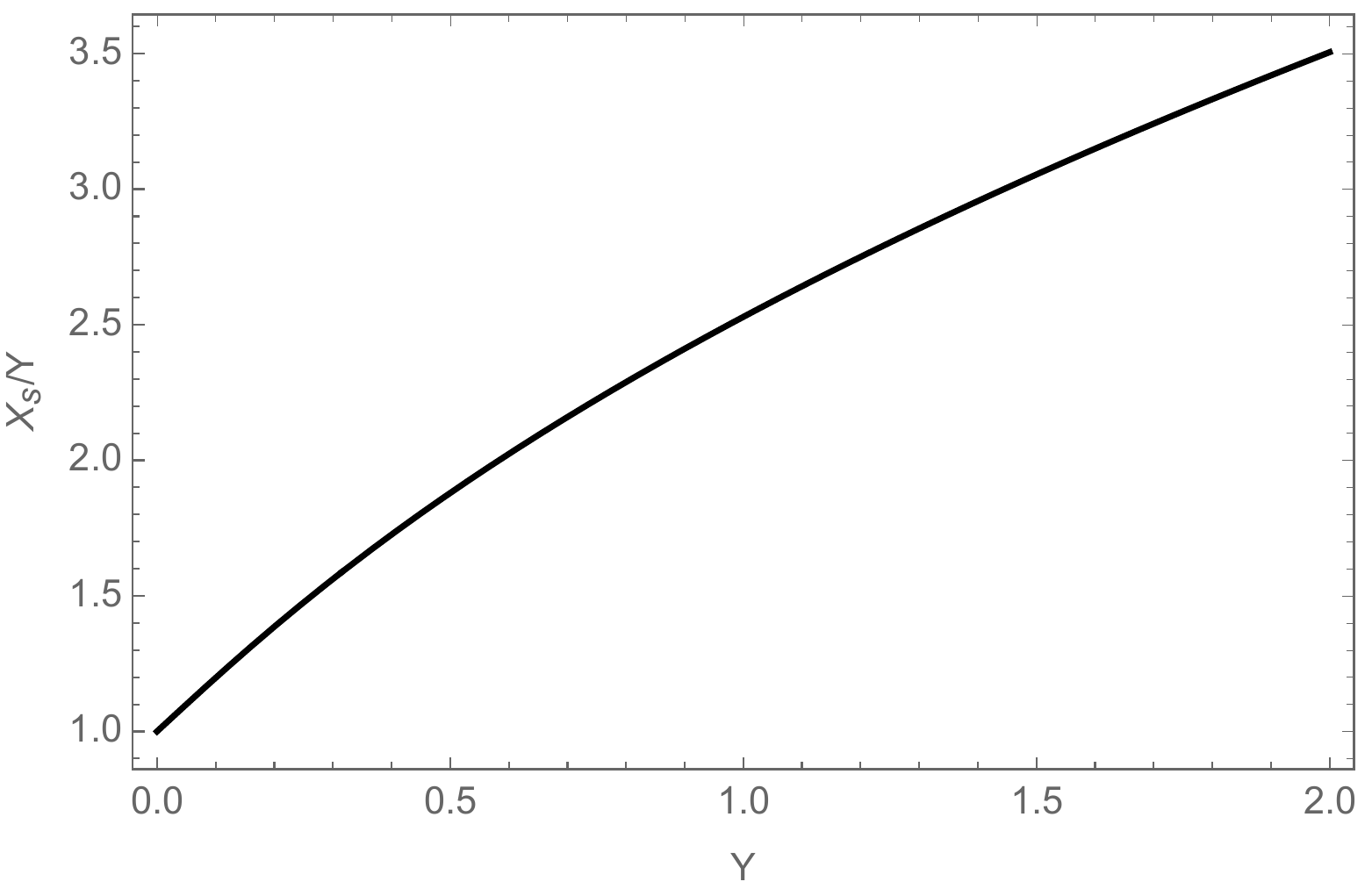}
\caption{Compactness in Eq.~\eqref{XcE} with $q_\Phi=1$.}
\label{XsY}
\end{figure}

For $n>1$, Eq.~\eqref{dbR} now reads
\be
Y
=
\frac{X}{\left(1+6\,q_\Phi\,X\right)^{1/3}}
\ ,
\ee
which can be used to determine the compactness $X$ from the behaviour of the density at the star surface.
Of the three solutions for $X$, only one is real and positive and reads 
\be
\frac{X_{\rm s}}{Y}
=
\frac{2^{1/3}\left(1+\sqrt{1-32\,q_\Phi^3\,Y^3}\right)^{2/3}+4\,q_\Phi\,Y}
{2^{2/3}\left(1+\sqrt{1-32\,q_\Phi^3\,Y^3}\right)^{1/3}}
\ ,
\label{XcE}
\ee
which holds for all values of $Y>0$.
We note that $X_s=Y$ for $q_\Phi=0$, corresponding to the Newtonian theory, and an expansion
for small $Y$ yields
\be
\frac{X_s}{Y}
\simeq
1
+2\,q_\Phi\,Y
+\frac{8}{3}\,q_\Phi^3\,Y^3
\ .
\ee 
This shows that the compactness is always larger in the bootstrapped theory than it would be in the
Newtonian case for the same $Y$ (see Fig.~\ref{XsY} for a plot of the exact result).
\par
Finally, we can determine more explicitly the boundary behaviour (at $r\lesssim R$) of the density,
which is governed by the equation
\be
\gamma\,n\,R\,\frac{\rho'}{\rho^{2-n}}
\simeq
-Y
\ ,
\ee
with the conditions that $\rho_R=0$ and $n>1$.
For $Y>0$, we then find
\be
\rho
&\!\!\simeq\!\!&
\left[
\frac{n-1}{n}\,
\left(
1
-
\frac{r}{R}
\right)
\frac{Y}{\gamma}
\right]^{1/(n-1)}
\nonumber
\\
&\!\!\simeq\!\!&
\left[
\frac{n-1}{n\,\gamma}\,
\left(
1
-
\frac{r}{R}
\right)
\frac{X}{(1+6\,q_\Phi\,X)^{1/3}}
\right]^{1/(n-1)}
\ .
\ee
Upon expanding for small compactness, one then obtains
\be
\rho
\simeq
\left[
\frac{n-1}{n\,\gamma}\,
\left(
1
-
\frac{r}{R}
\right)
X
\right]^{1/(n-1)}
\left(
1-\frac{2\,q_\Phi\,X}{n-1}
\right)
\ ,
\ee
which shows that the density near the surface must be smaller in the bootstrapped theory
than it is in the Newtonian theory for a star of given (small) compactness.
We remark once more that this result is independent of the non-relativistic limit $q_c\to 0$
because relativistic corrections proportional to $q_c$ are subleading near the surface of the
source for $n>1$.
\par
To summarise, we have obtained the general behaviour of the density required by the equation of
state~\eqref{eosB} to be compatible with a smooth potential across the surface.
We remark that one could relax the condition~\eqref{dbR} on the derivative of the potential in order to allow for
a (vanishingly thin) solid crust.
In Section~\ref{S:gaussian}, we shall employ an analytical approximation for the density inside the whole star and
consider in particular the text-book equations of state for relativistic and non-relativistic fermions,
so that $1<n<2$ and $\rho'(R)=\rho(R)=0$.
Before that, we will tackle the problem numerically.
\subsection{Density equation and numerical solutions}
\label{S:numerical}
For the cases of interest, the potential is expressed exactly in terms of the density by Eq.~\eqref{VR_rho}
with $1<n<2$ and $\beta$ given in Eq.~\eqref{beta_0}.
This makes it more convenient to rewrite the second order ordinary differential equation~\eqref{EOMV}
inside the source as an equation for the density $\rho=\rho(r)$, with the two boundary conditions
\be
\rho(R)
&\!\!=\!\!&
0
\label{rho1}
\\
\rho'(0)
&\!\!=\!\!&
0
\ .
\label{drho0}
\ee
Furthermore, in order to solve for the density numerically, we shall introduce dimensionless
variables by using $R$ as the unit of length.
For instance, we write the radial coordinate $r=R\,\bar r$ and note that the compactness $X$ is already 
dimensionless.
Likewise, the dimensionless density $\bar\rho$ is defined by
\be
\gn\,\rho
=
\frac{X\,\bar\rho}{R^2}
\ee
and the polytropic equation of state~\eqref{eosB} yields
\be
\gn\,p
=
\frac{\bar\gamma}{R^2}\left(X\,\bar\rho\right)^n
\ ,
\label{eosD}
\ee
where the dimensionless $\bar\gamma$ should not be confused with $\tilde\gamma$.~\footnote{We
note that $\rho_0$ is not a convenient parameter here since the central density cannot be set freely.}
The equation for the dimensionless density then reads
\be
&&
\strut\displaystyle
\frac{\partial_{\bar r}}{\bar r^2}
\left[
\frac{n\,\bar\gamma\,\bar r^2\,\partial_{\bar r}(\ln\bar\rho)}
{q_c\,\bar\gamma+(X\,\bar\rho)^{1-n}}
\right]
=
-4\,\pi\,X\,\bar\rho
\left[1+q_c\,\bar\gamma\,(X\,\bar\rho)^{n-1}\right]
\nonumber
\\
&&
\strut\displaystyle
-
\frac{2\,q_\Phi\,n^2\,(n-1)\,\bar\gamma^2\,[\partial_{\bar r}(\ln\bar\rho)]^2}
{\left[q_c\,\bar\gamma+(X\,\bar\rho)^{1-n}\right]^2
\left\{(n-1)\left(1-4\,q_\Phi\,V_R\right)
+4\,q_\Phi\,n\,q_c^{-1}\ln \left[1+q_c\,\bar\gamma\,(X\,\bar\rho)^{n-1} \right]
\right\}}
\ ,
\label{eomRho}
\ee
in which we note that $V_R$ is given in Eq.~\eqref{VR} and is a function of $X$ only.
\par
\begin{figure}[t]
\centering
\includegraphics[width=8cm]{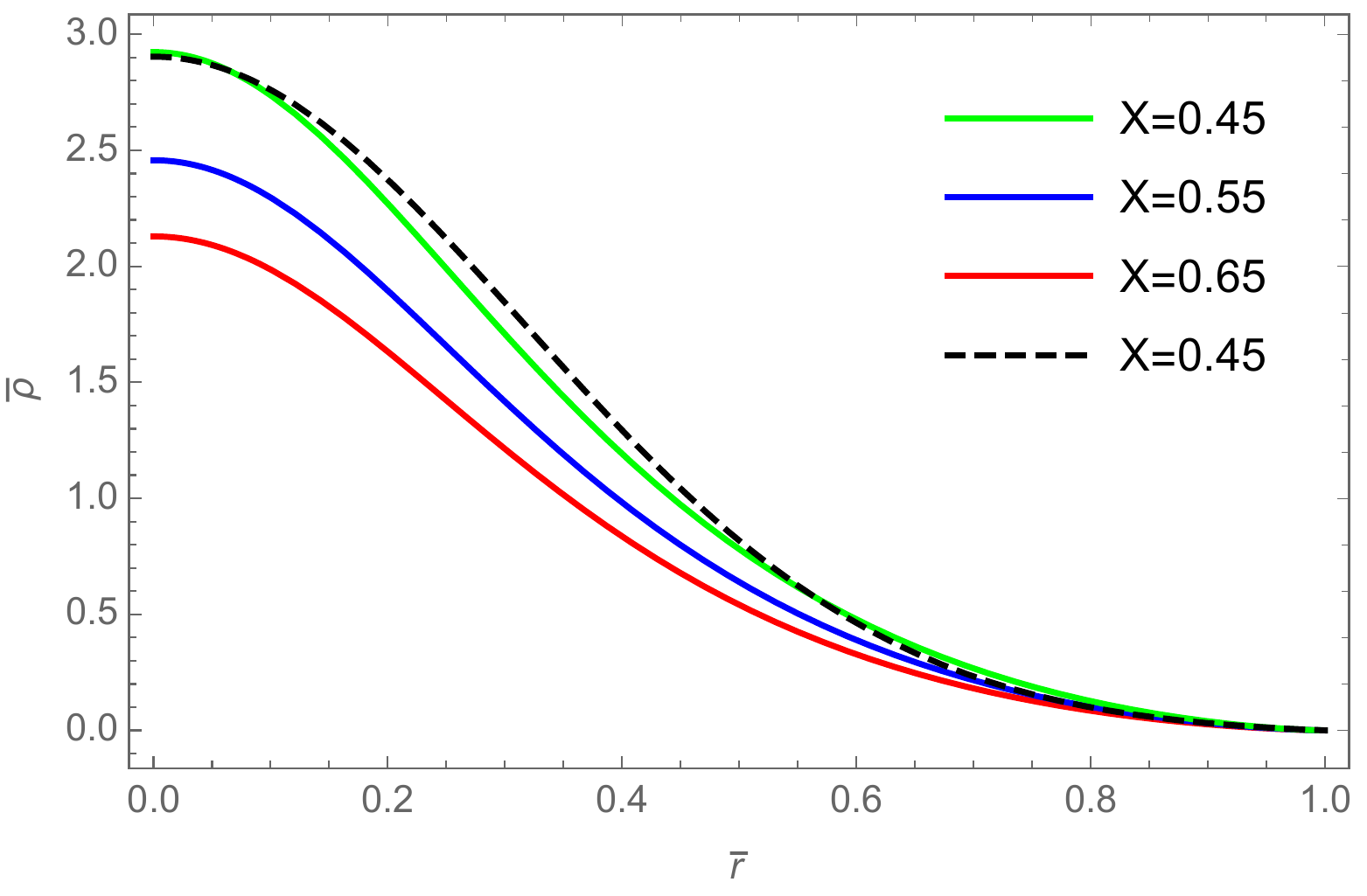}
$\ $
\includegraphics[width=8cm]{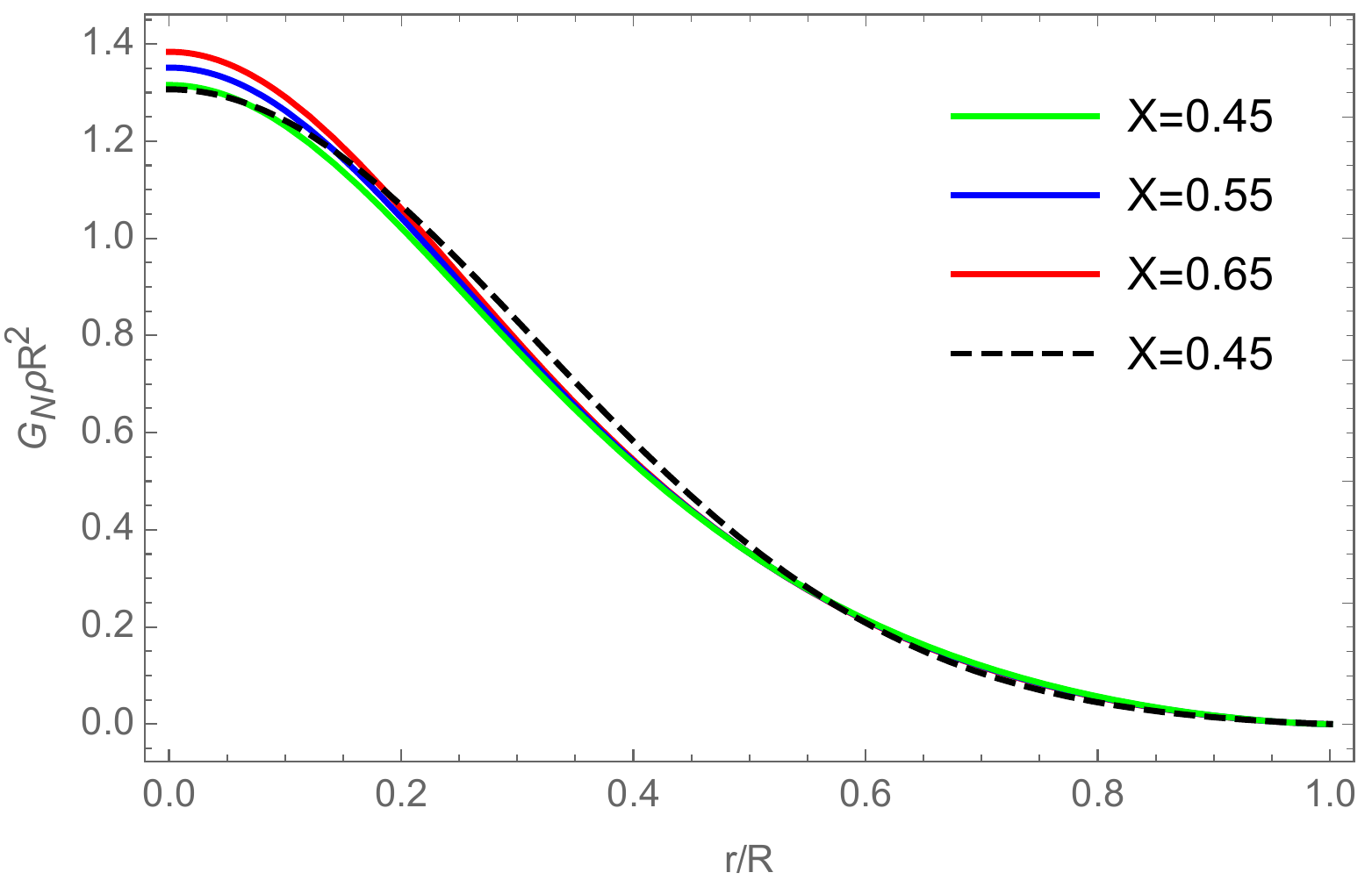}
\\
$\ $
\\
\includegraphics[width=5.3cm,height=4cm]{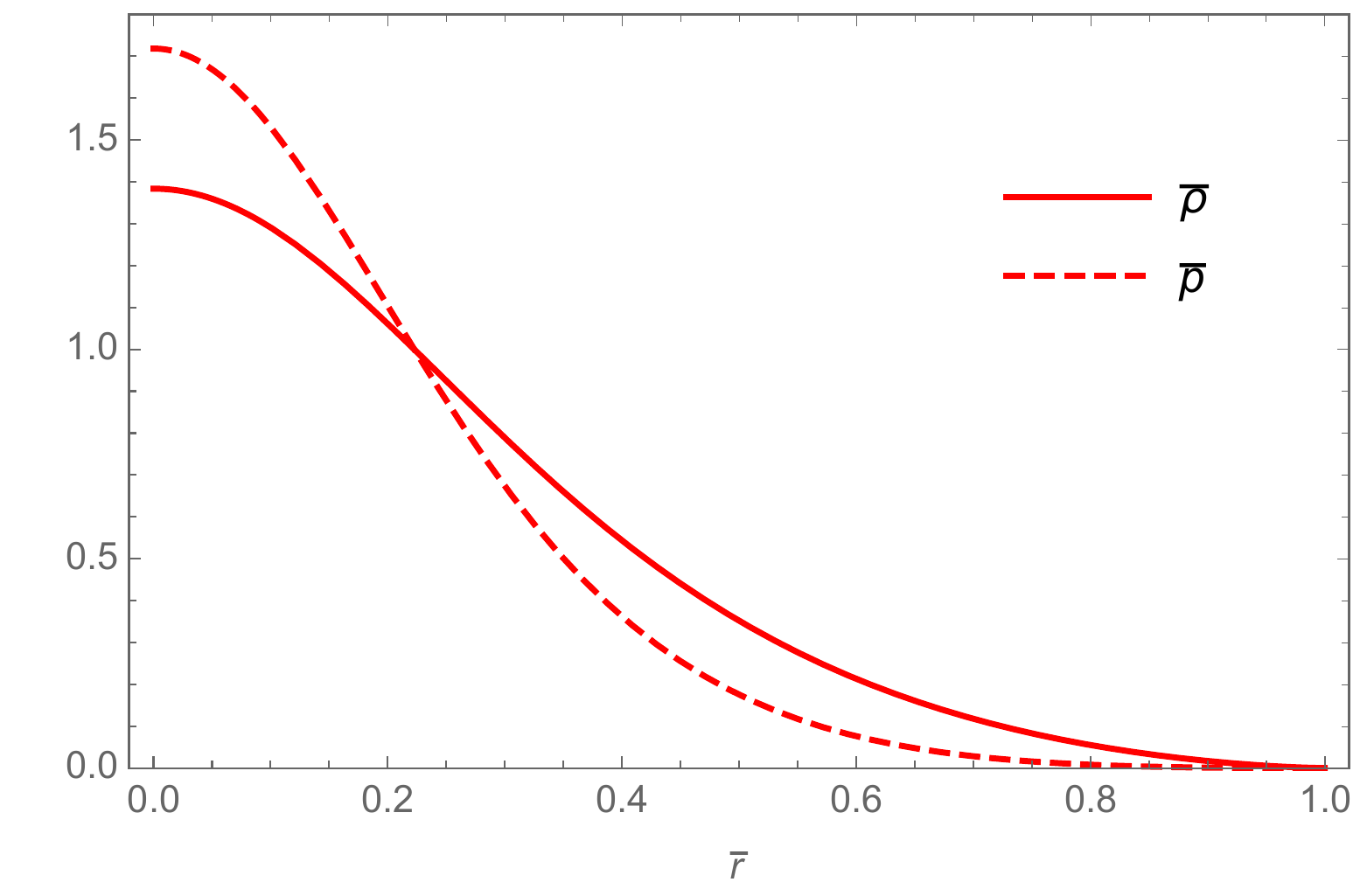}
\includegraphics[width=5.3cm,height=4cm]{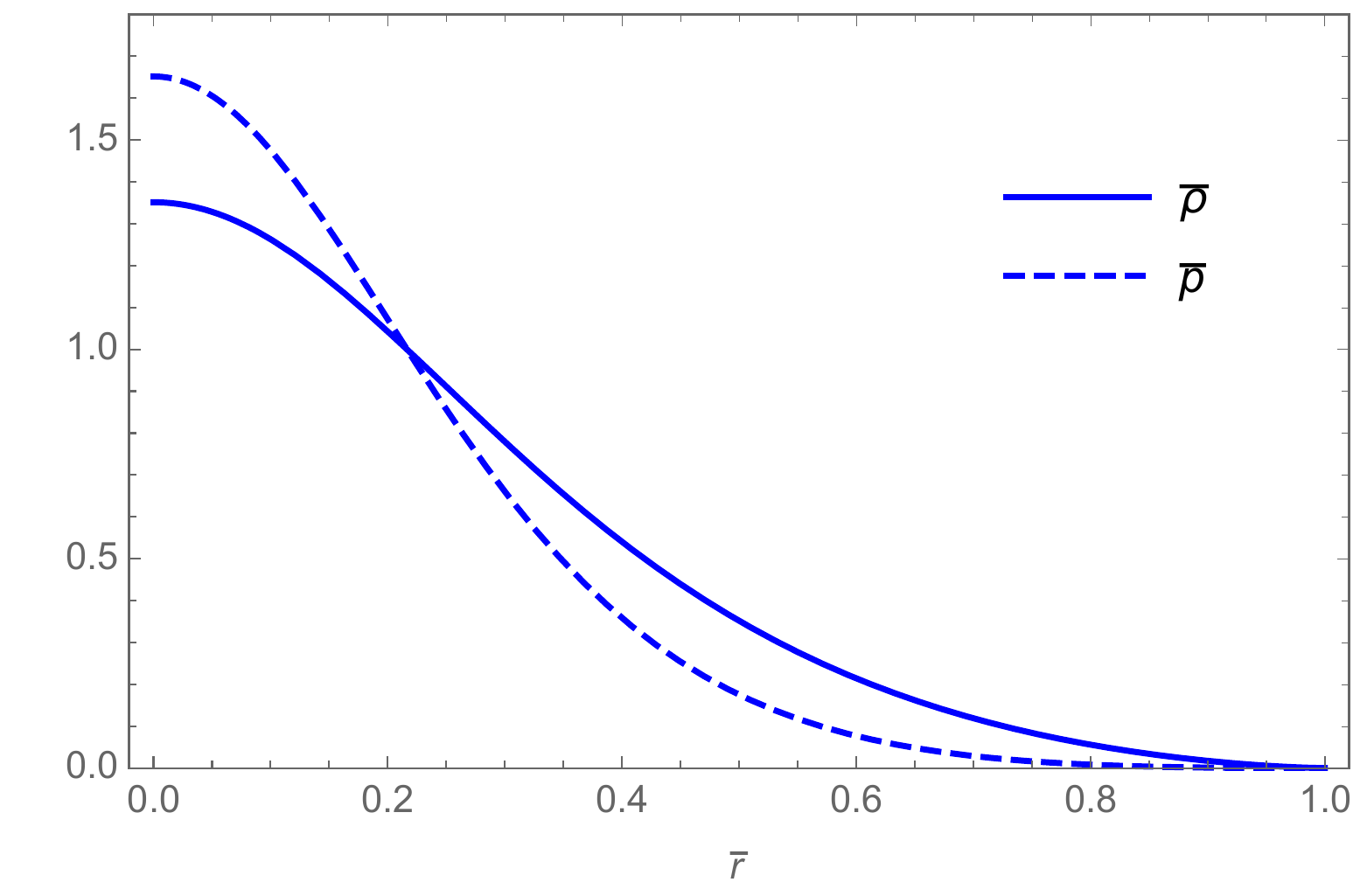}
\includegraphics[width=5.3cm,height=4cm]{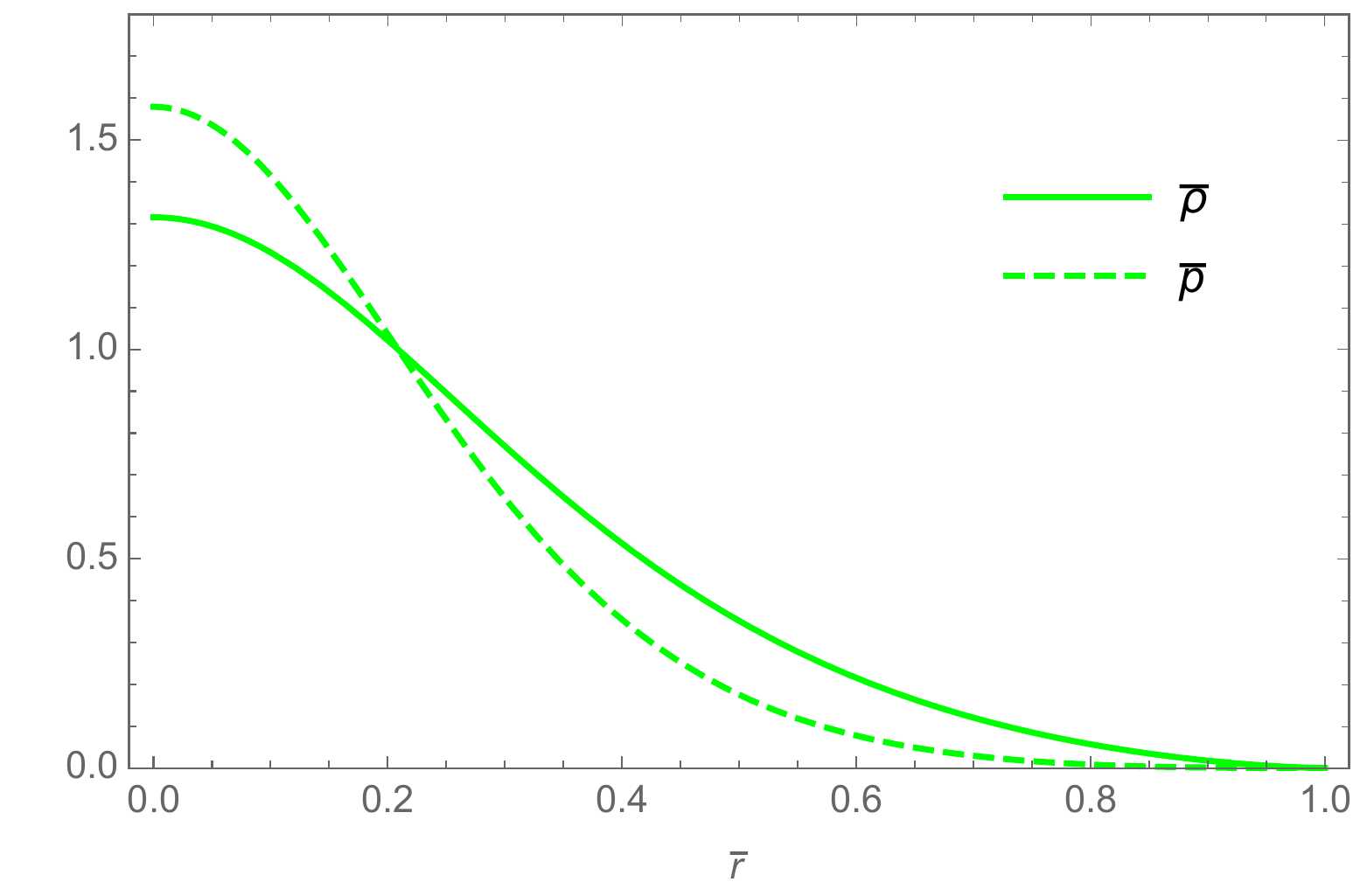}
\caption{Upper panels:
density profile calculated numerically for $\tilde\gamma=1$, $n=5/3$ (solid lines)
and Gaussian approximation (dashed line) for the lowest compactness
(left panel: dimensionless quantities; right panel: dimensionful quantities).
Lower panels:
density (solid lines) and pressure (dashed lines) for the cases in the upper panels}
\label{RhoN5s3}
\end{figure}
\begin{figure}[h]
\centering
\includegraphics[width=8cm]{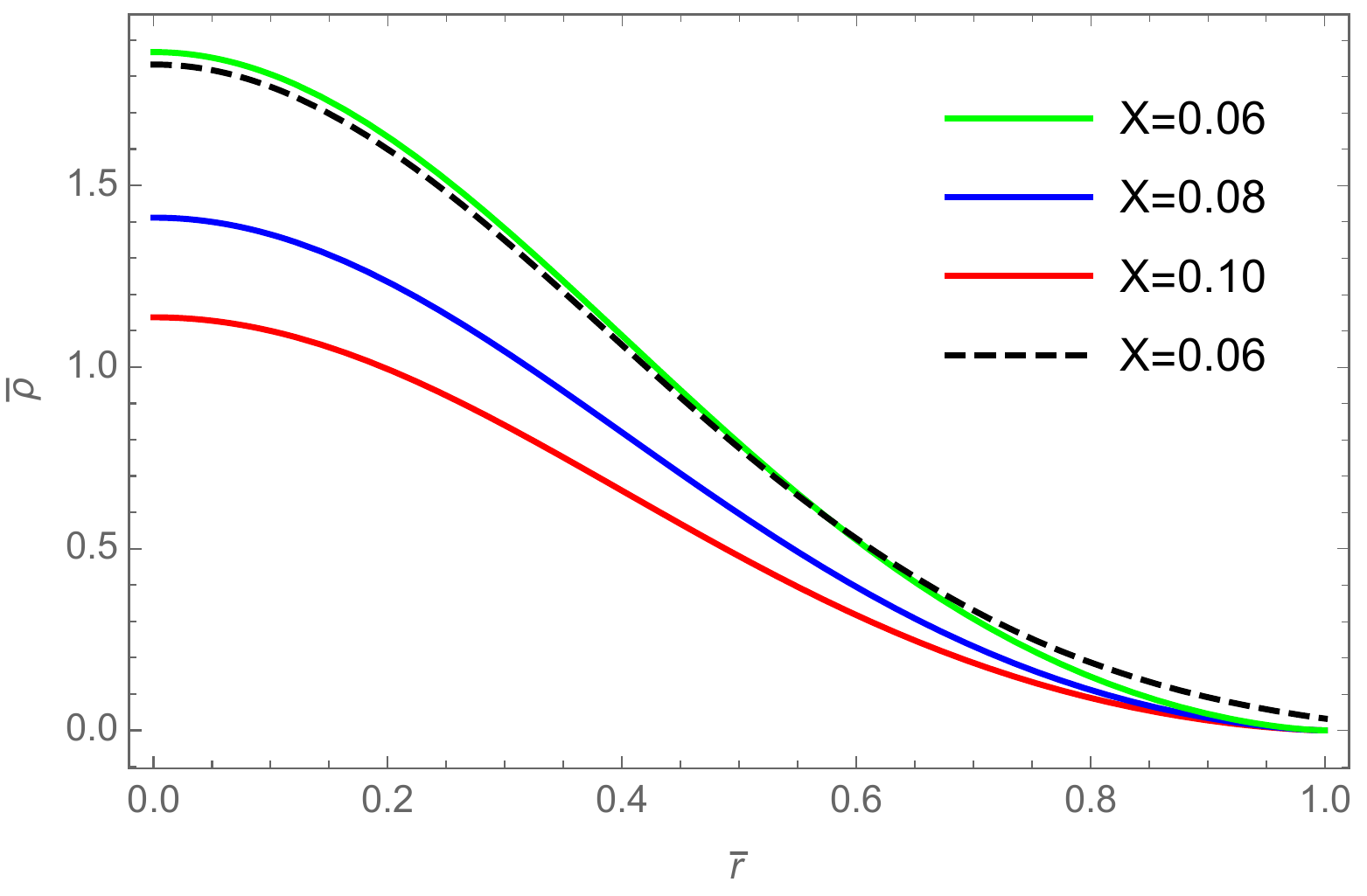}
$\ $
\includegraphics[width=8cm]{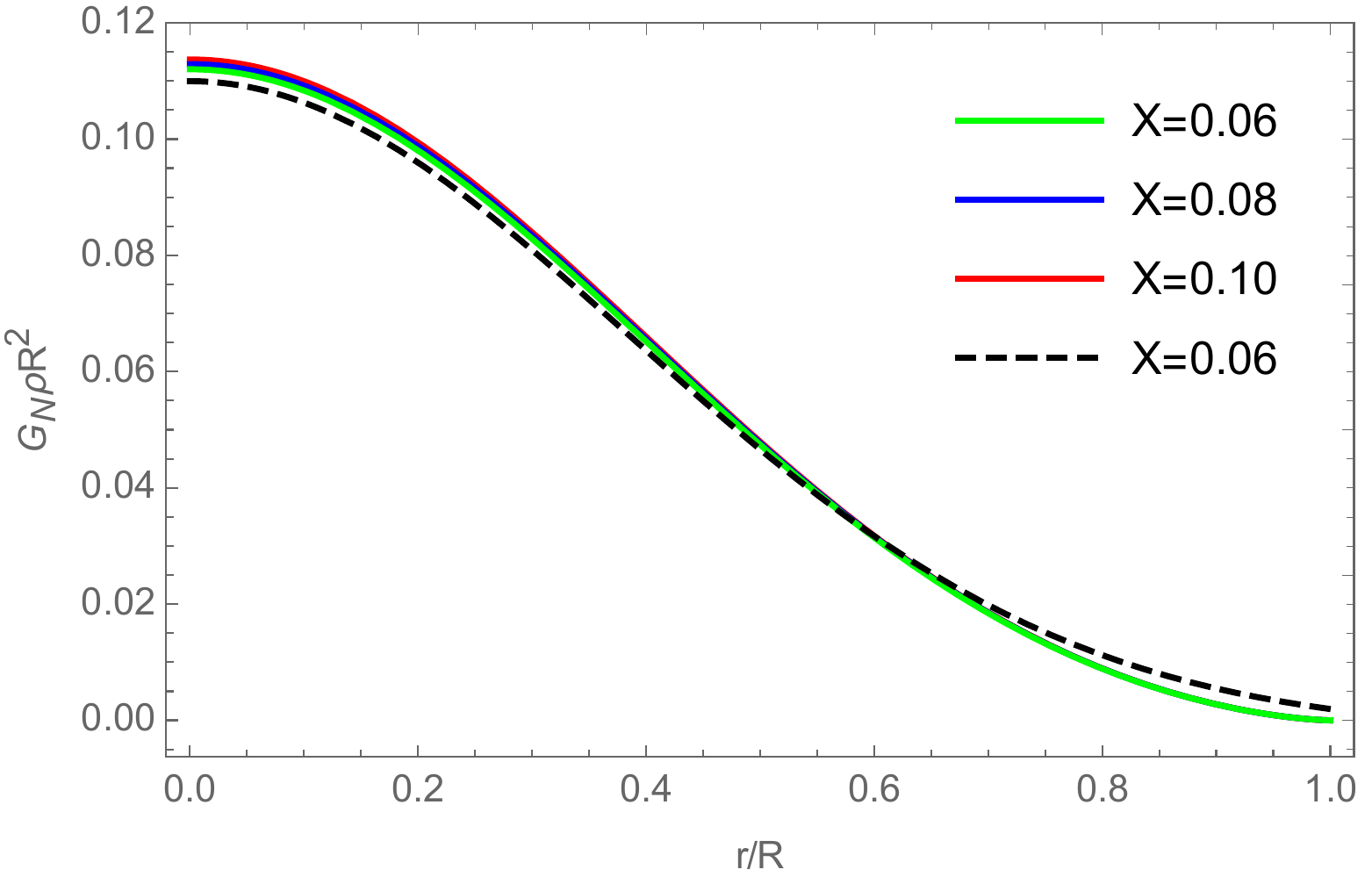}
\\
$\ $
\\
\includegraphics[width=5.3cm,height=4cm]{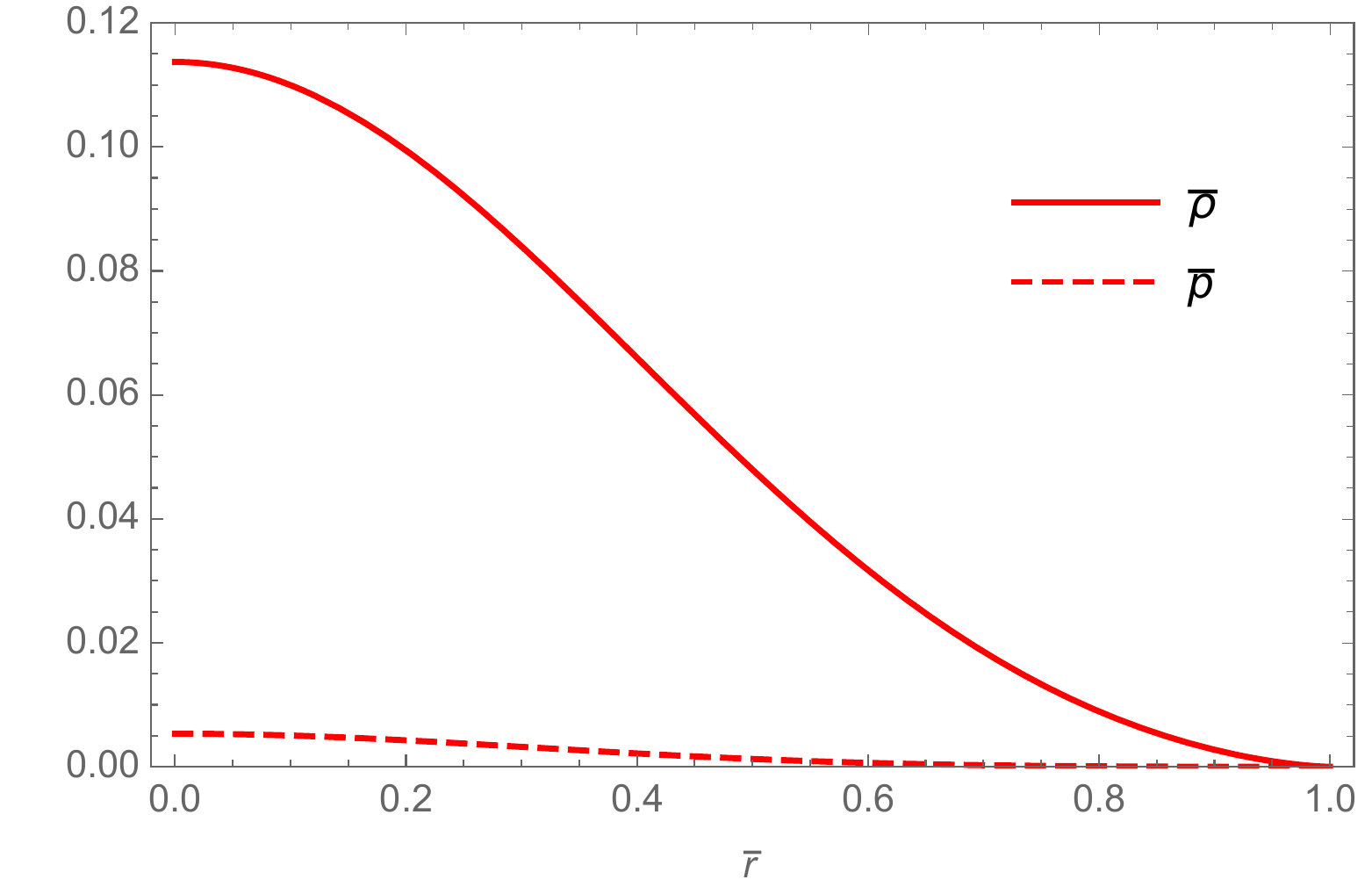}
\includegraphics[width=5.3cm,height=4cm]{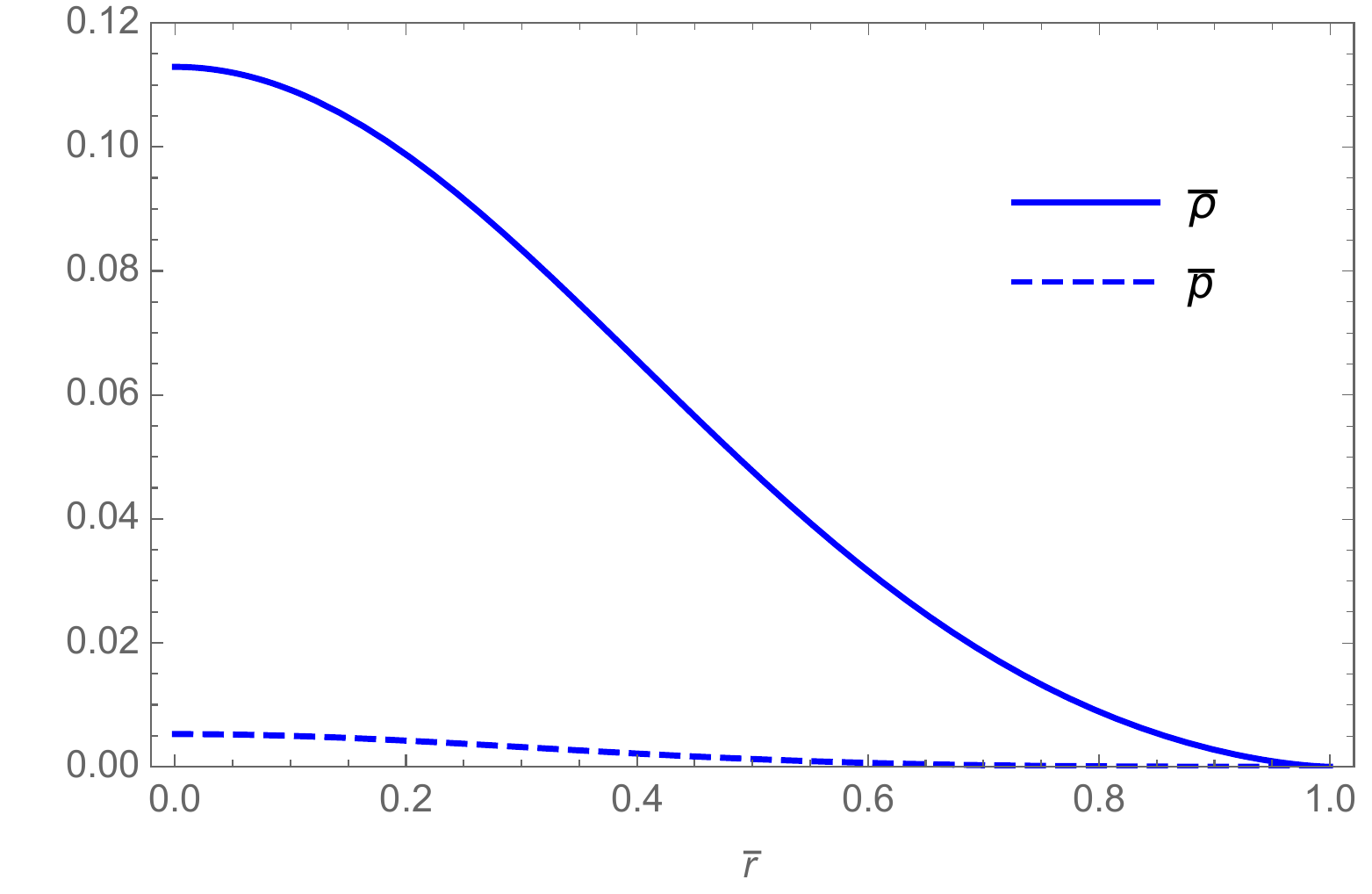}
\includegraphics[width=5.3cm,height=4cm]{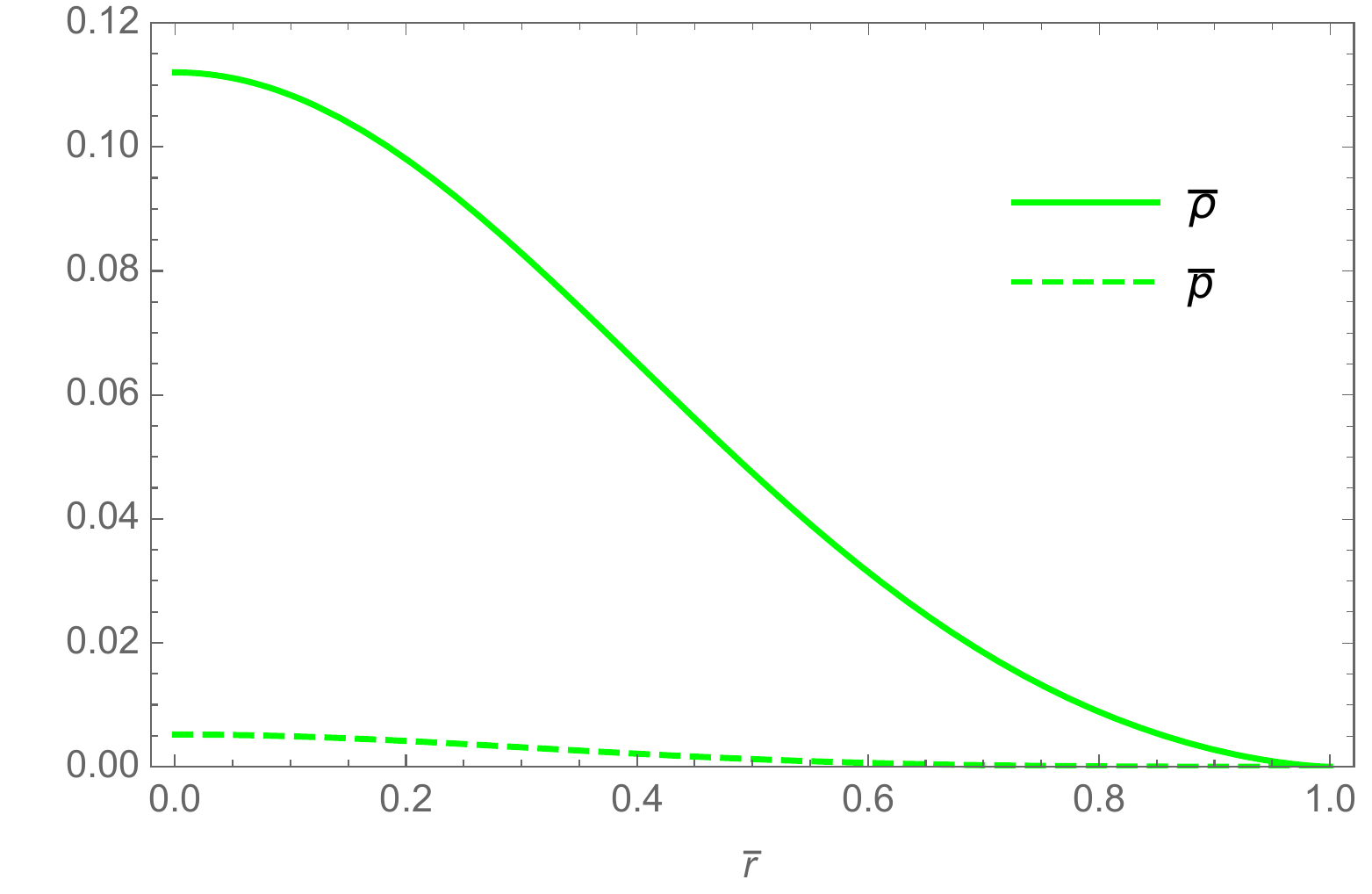}
\caption{Upper panels:
density profile calculated numerically for $\tilde\gamma=0.2$, $n=5/3$ (solid lines)
and Gaussian approximation (dashed line) for the lowest compactness (left panel: dimensionless quantities;
right panel: dimensionful quantities).
Lower panels:
density (solid lines) and pressure (dashed lines) for the cases in the upper panels}
\label{RhoN5s3b}
\end{figure}
\begin{figure}[h]
\centering
\includegraphics[width=8cm]{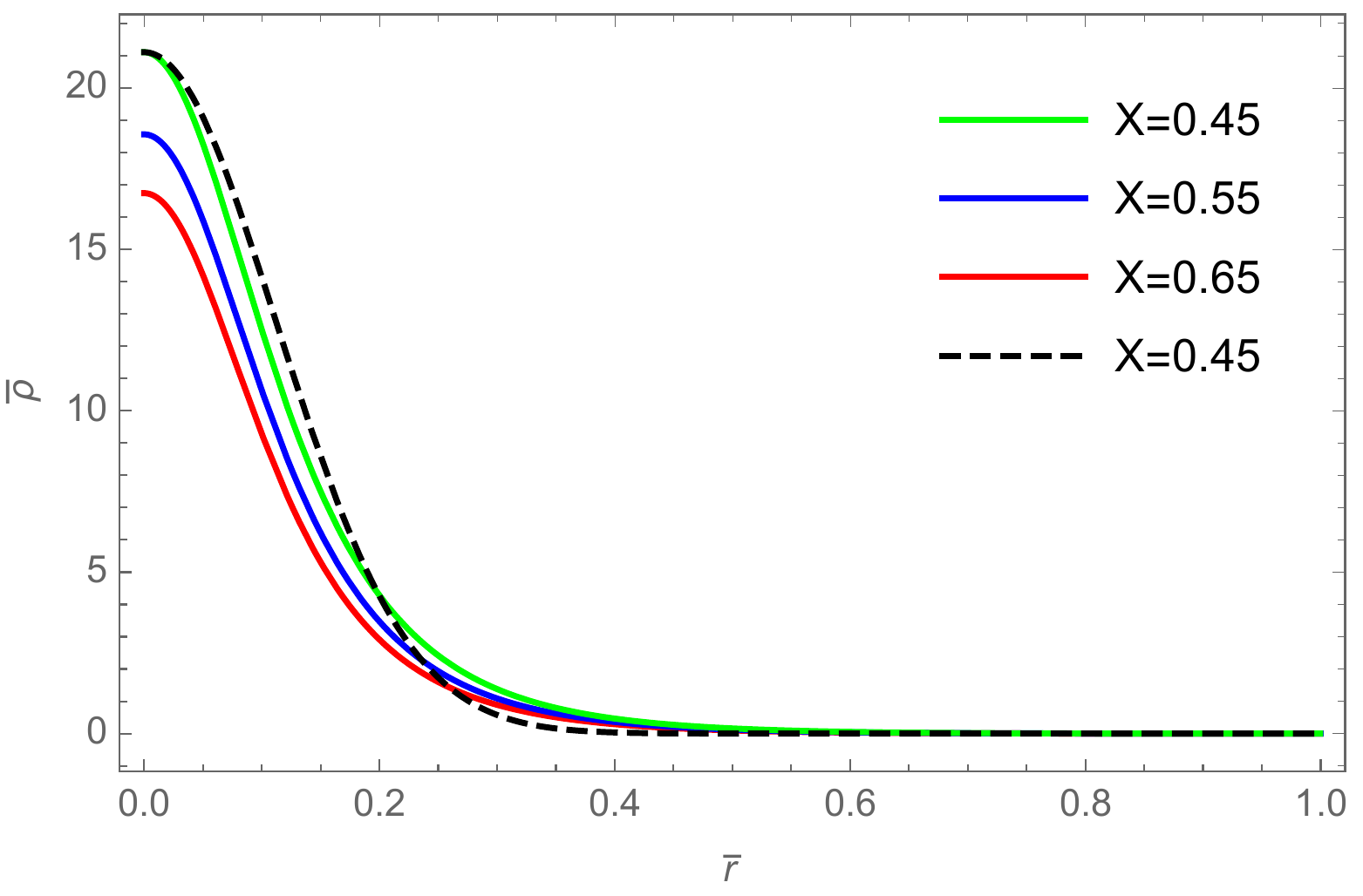}
$\ $
\includegraphics[width=8cm]{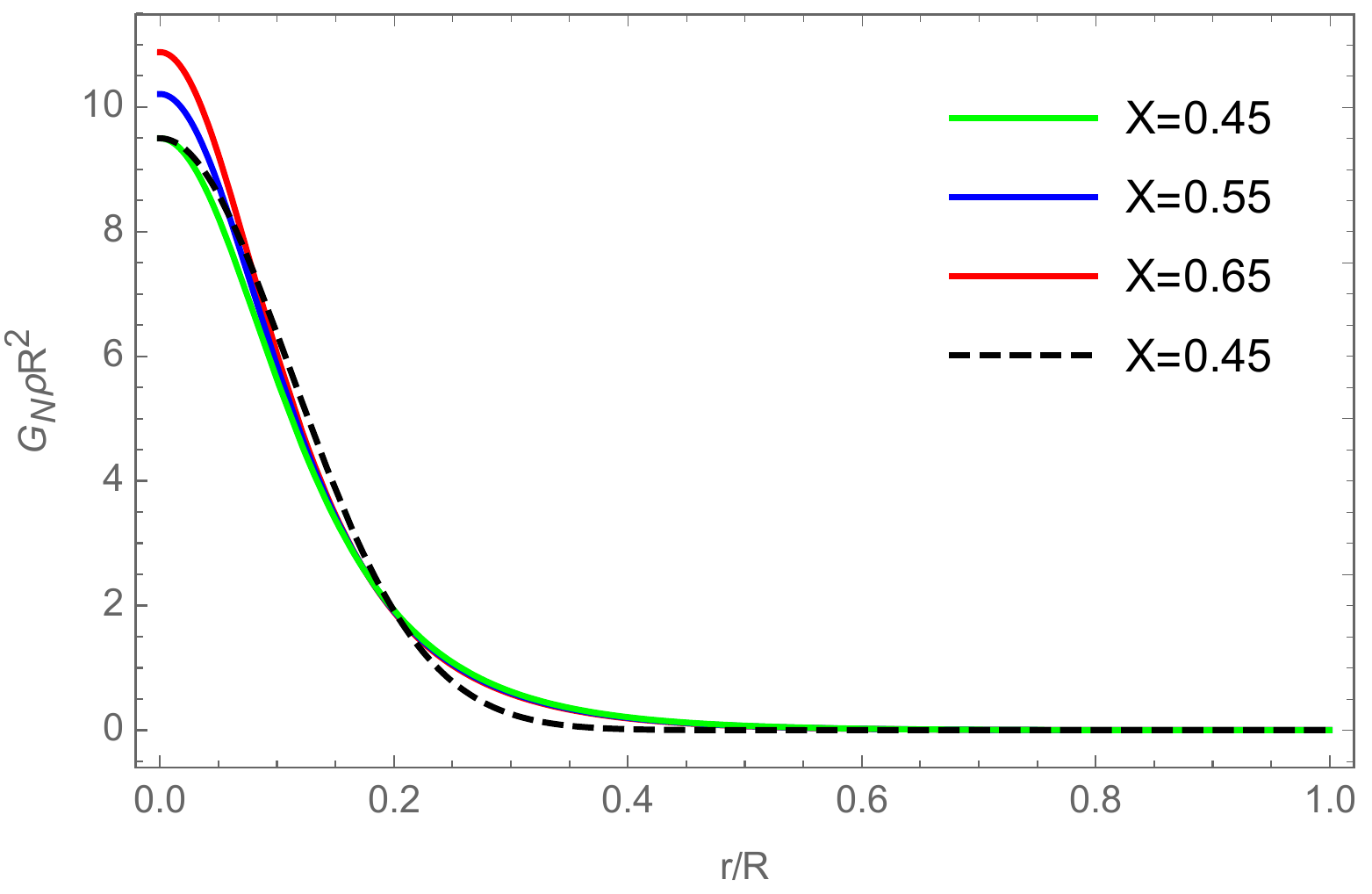}
\\
$\ $
\\
\includegraphics[width=5.3cm,height=4cm]{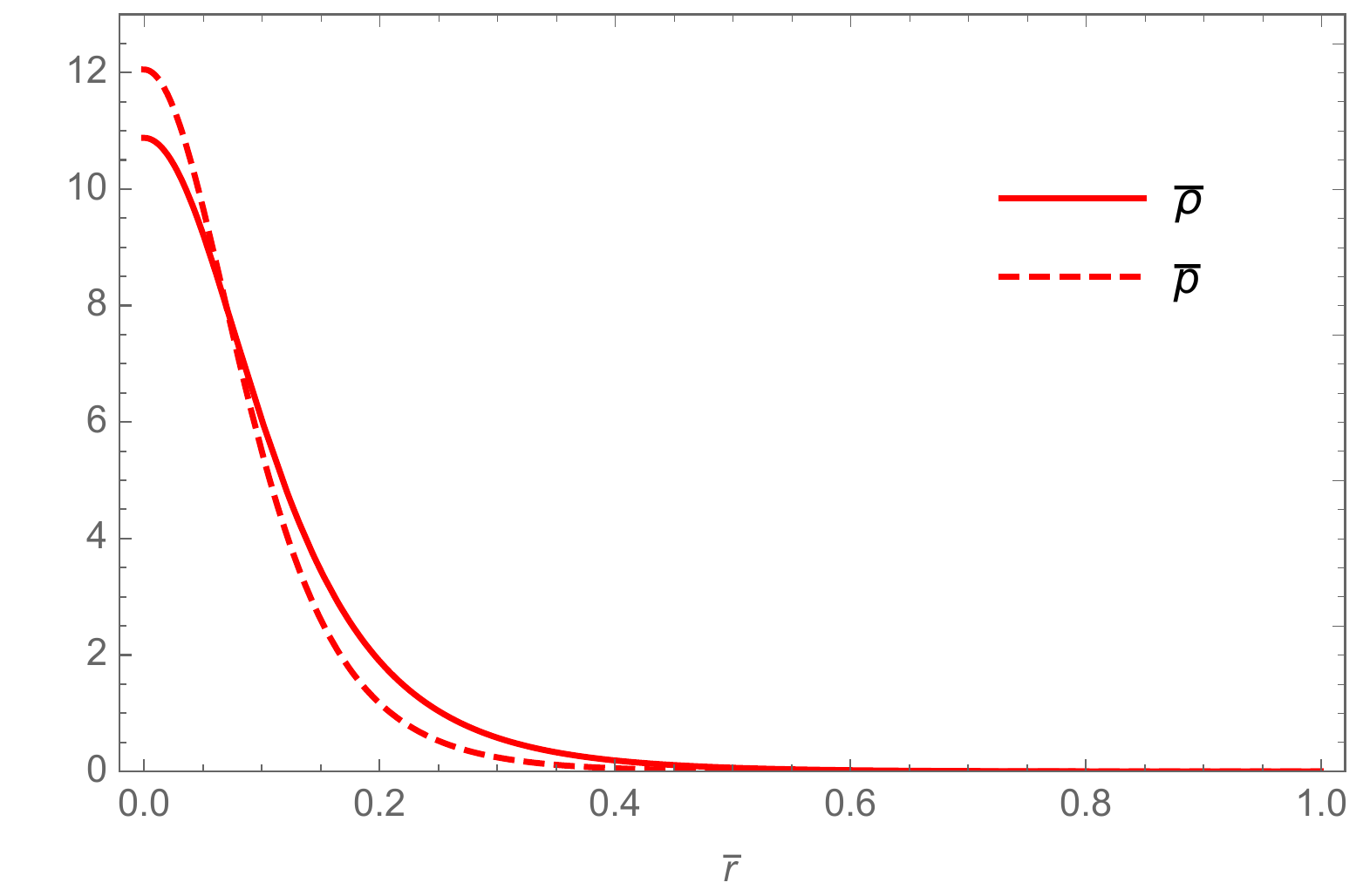}
\includegraphics[width=5.3cm,height=4cm]{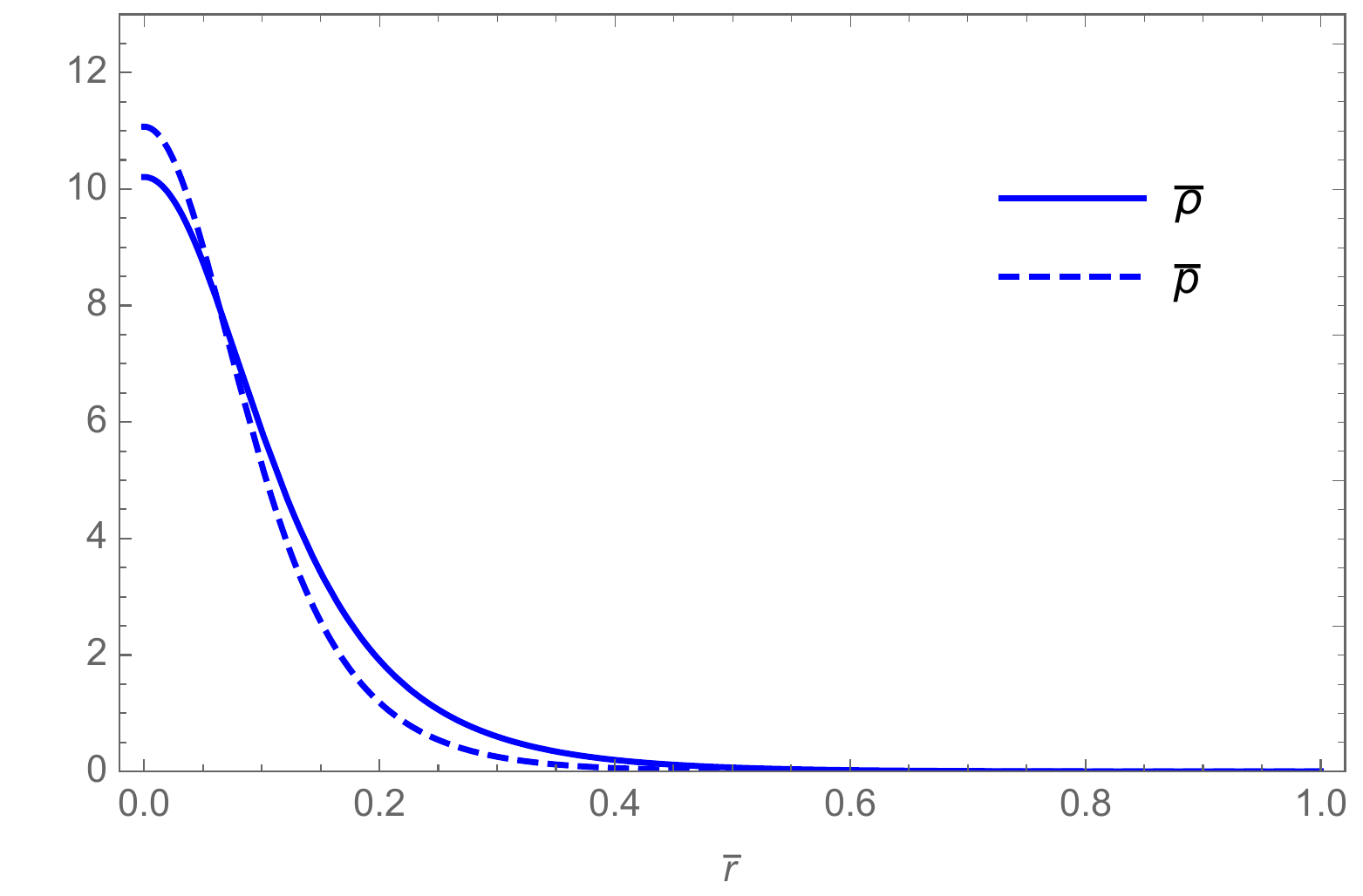}
\includegraphics[width=5.3cm,height=4cm]{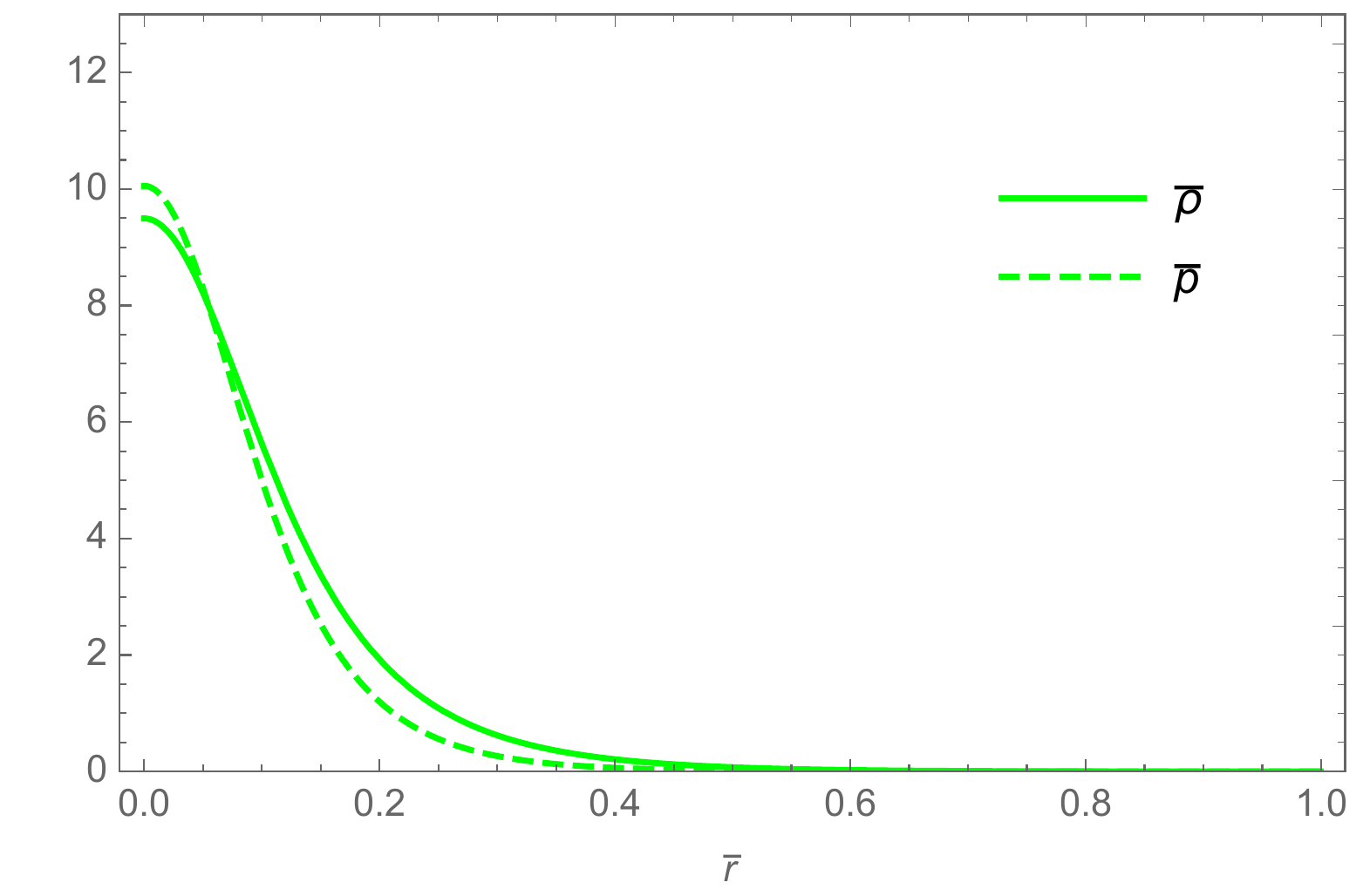}

\caption{Upper panels:
density profile calculated numerically for $\tilde\gamma=0.5$, $n=4/3$ (solid lines)
and Gaussian approximation (dashed line) for the lowest compactness (left panel: dimensionless quantities;
right panel: dimensionful quantities).
Lower panels:
density (solid lines) and pressure (dashed lines) for the cases in the upper panels}
\label{RhoN4s3}
\end{figure}

We have performed a preliminary numerical analysis of the above equation and boundary conditions for $q_c=q_\Phi=1$.
We also want to avoid values of the compactness corresponding to Newtonian black holes corresponding to
$X\gtrsim 0.7$ (as discussed in Section~\ref{S:vacuum}).
The relevant parameter space is thus given by $0<X<0.7$, $1<n<2$ and $\bar\gamma>0$,
a complete analysis of which would require extensive numerical works beyond our present scope.
A first interesting result is that, for fixed values of $n$ and $\bar\gamma$, solutions only exist for certain ranges
of $X$, similarly to the case of General Relativity.
Examples of high compactness are given in Figs.~\ref{RhoN5s3} for $n=5/3$,
from which we see that the larger $X$, the flatter the dimensionless profile of $\bar\rho$,
whereas the fully dimensional density grows larger in the centre and so does the pressure.
Another preliminary result is that solutions are found for lower values of $X$ only by suitably lowering
$\bar\gamma$ correspondingly.
Examples for the same $n=5/3$ are given in Figs.~\ref{RhoN5s3b}.
Finally, we have found that smaller values of $n$ produce more peaked profiles, like is shown for $n=4/3$
in Fig.~\ref{RhoN4s3}.
\par
In all the case we have been able to solve Eq.~\eqref{eomRho}, the density profile can be rather closely
approximated with a Gaussian function.
In the following, we shall therefore take the opposite perspective and try to determine the
polytropic parameters compatible with given Gaussian profiles. 
\section{Gaussian density profiles}
\label{S:gaussian}
\setcounter{equation}{0}
We start from assuming the density profile of the self-gravitating object is given by
\be
\rho
=
\left\{
\begin{array}{lr}
\strut\displaystyle
\rho_0\,e^{-\frac{r^2}{b^2\,R^2}} 
\ ,
&
r\leq R
\\
\\
\strut\displaystyle
0 \,
\ ,
&
r>R
\ .
\end{array}
\right.
\label{rho_gaussian}
\ee
Since $\rho_R\equiv \rho(R)>0$, the density~\eqref{rho_gaussian} contains a step-like discontinuity at $r=R$,
like the uniform profiles analysed in Refs.~\cite{Casadio:2019cux,BootN}.
Of course, such a discontinuity is incompatible with a polytropic equation of state if one continues to require
vanishing pressure at the surface.
However, we can set the central density $\rho_0$ and the width $b$ such that
\be
-\lim_{r\to R}
\left(\frac{\rho'}{\rho^{2-n}}\right)
=
\frac{\rho_0^{n-1}}{b^2}\,e^{-(n-1)/b^2}
=
\frac{Y}{2\,\gamma}
\ ,
\ee
and note that for $b\ll 1$, we can have $\rho_R\ll \rho_0$.
A mild discontinuity of this form could be tolerable, for instance, by assuming that the surface of the object
is covered by a thin solid crust with a tension that balances the non-vanishing pressure. 
\par
Technically, the discontinuity could also be removed completely by subtracting the constant $\rho_R$ from the
profile~\eqref{rho_gaussian} for $r\le R$.
In so doing, one would however introduce more serious obstacles with the continuity of the first derivative
of the potential, since the denominator in Eq.~\eqref{VR_rho1} would then vanish and $V'$ correspondingly diverge.
We therefore find it still preferable to allow for a (slight) discontinuity at the surface with $\rho_R\ll \rho_0$.
It needs to be mentioned that, due to this discontinuity, Eq.~\eqref{beta_0} for $\beta$ also gets modified,
as will be seen later. 
\par
Starting from Eqs.~\eqref{VR_rho1} and~\eqref{VR_rho}, and using the two boundary conditions~\eqref{bR}
and~\eqref{dbR}, we can determine both $\beta$ and $\gamma$ as functions of the other remaining parameters.
Continuity of the first derivative of the potential~\eqref{dbR} allows us to express one of the parameters
which determine the equation of state, $\tilde\gamma$ (respectivelly $\gamma$), in terms of the polytropic
index $n$, the compactness $X$ and the width $b$, as
\be
\tilde\gamma
=
\frac{X\,e^{(n-1)/b^2}}{\frac{2\,n}{b^2}\left(1+6\,q_\Phi\,X\right)^{1/3}-q_c\,X}
\ .
\ee
Using the boundary condition~\eqref{bR}, we then determine 
\be
\beta
&\!\!=\!\!&
V_R
+
\frac{n}{(n-1)\,q_c}\,\ln \left[1+q_c\,\tilde\gamma\,e^{(1-n)/b^2} \right] \nonumber\\
&\!\!=\!\!&
V_R
+
\frac{n}{(n-1)\,q_c}\,\ln \left[1+\frac{q_c\,X}{\frac{2\,n}{b^2}\left(1+6\,q_\Phi\,X\right)^{1/3}-q_c\,X} \right]
\ .
\ee
\par
These expressions for $\tilde\gamma$ and $\beta$ can be substituted into the field equation~\eqref{EOMV},
but no exact analytical expression can then be found for the remaining parameters.
We can proceed by expanding both sides of Eq.~\eqref{EOMV} in power series around $r=0$.
Equating the lowest order terms yields
\be
\rho_0
&\!\!=\!\!&
\frac{3\,n\,\tilde\gamma}{2\,\pi\,b^2\,\gn\,R^2\left(1+q_c\,\tilde\gamma\right)^2}
\nonumber
\\
&\!\!=\!\!&
\frac{3\,n\,X\,e^{(n+1)/b^2}\left[\frac{2\,n}{b^2}\left(1+6\,q_\Phi\,X\right)^{1/3}-q_c\,X\right]}
{2\,\pi\,\,b^2\,\gn\,R^2\left[e^{n/b^2}\,q_c\,X+e^{1/b^2}\left(\frac{2\,n}{b^2}\left(1+6\,q_\Phi\,X\right)^{1/3}-q_c\,X\right)\right]^2}
\ ,
\ee 
from which we can write
\be
\rho
&\!\!=\!\!&
\frac{3\,n\,\tilde\gamma\,e^{-r^2/b^2\,R^2}}{2\,\pi\,b^2\,\gn\,R^2\left(1+q_c\,\tilde\gamma\right)^2}
\nonumber
\\
&\!\!=\!\!&
\frac{3\,n\,X\,e^{(n+1-r^2/R^2)/b^2}\left[\frac{2\,n}{b^2}\left(1+6\,q_\Phi\,X\right)^{1/3}-q_c\,X\right]}
{2\,\pi\,\,b^2\,\gn\,R^2\left[e^{n/b^2}\,q_c\,X+e^{1/b^2}\left(\frac{2\,n}{b^2}\left(1+6\,q_\Phi\,X\right)^{1/3}-q_c\,X\right)\right]^2}
\ .
\ee 
\par
\begin{figure}[t]
\centering
\includegraphics[width=10cm]{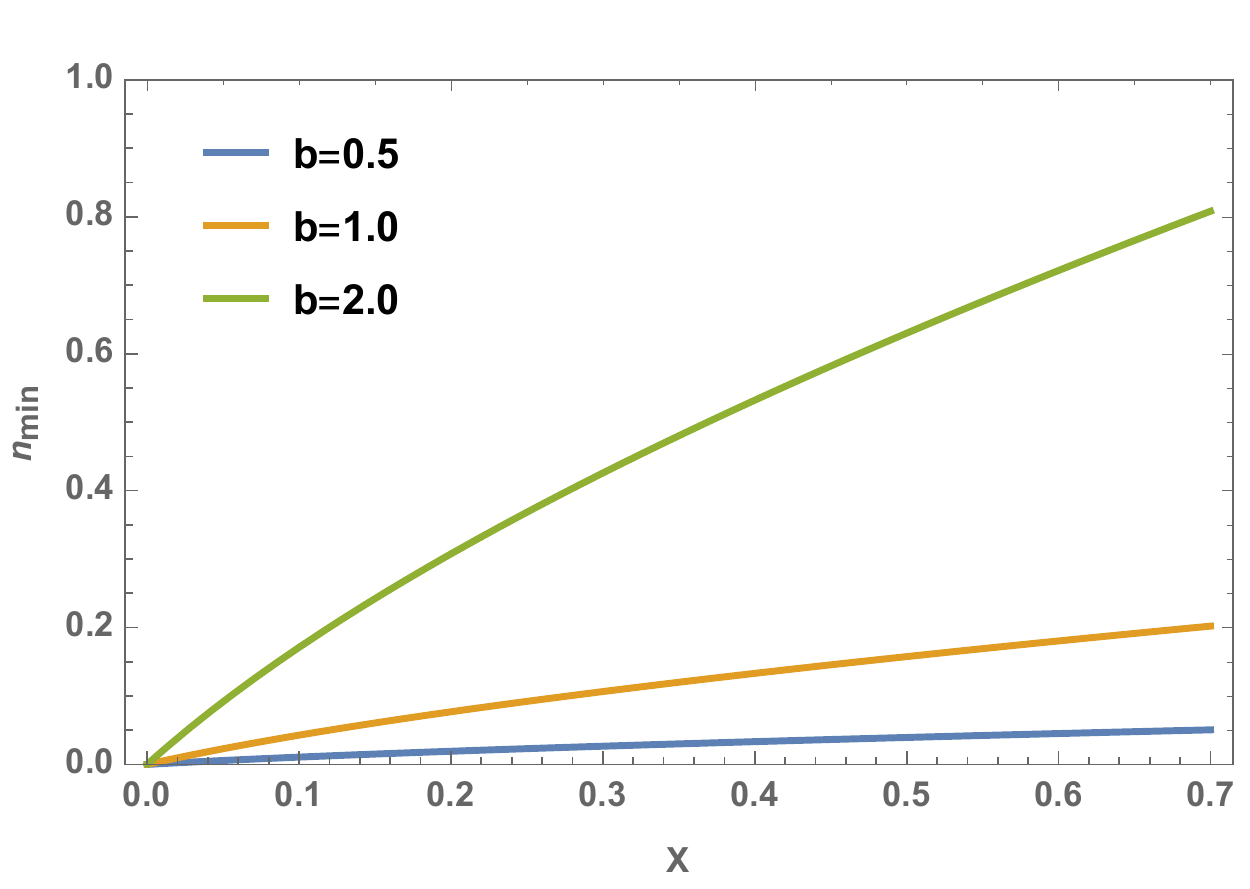}
\caption{Minimum value of the polytropic index $n$ for $q_c=q_\Phi=1$.}
\label{b_max_plot}
\end{figure}

We also notice that both $\rho_0$ and $\tilde\gamma$ must be positive, which holds if 
\be
n > \frac{q_c\,b^2\,X}{2\left(1+6\,q_\Phi\,X\right)^{1/3}}
\equiv
n_{\rm min}
\ .
\label{b_bound}
\ee
This is a non-trivial lower bound for the polytropic index depending on the compactness and width of the density profile.
Since we expect $1<n<2$, the compactness and width must satisfy
\be
8\left(1+6\,q_\Phi\,X\right)
<
q_c^3\,b^6\,X^3
<
64\left(1+6\,q_\Phi\,X\right)
\ ,
\ee
otherwise no $n$ exists and the star cannot be described by polytropic matter. 
The range of values for the compactness $X$ that will be considered further covers all possible types
of sources, from very low densities to objects on the brink of contracting behind the event horizon and
becoming black holes, as discussed in Section~\ref{S:vacuum}. 
The lower bound $n_{\rm min}$ from Eq.~\eqref{b_bound} is shown in Fig.~\ref{b_max_plot} for $q_c=q_\Phi=1$
and $0<X<0.7$.
The entire range $1<n<2$ is clearly allowed.
\par
The ADM mass $M$ and the proper mass $M_0$ of the star are generally different.
The proper mass is obtained from the volume integral of the density in Eq.~\eqref{proper_mass},
which, for our Gaussian distributions, reads
\be
M_0
=
\frac{3\,n\,X\,R\,e^{n/b^2}\left[\frac{2\,n}{b^2}\left(1+6\,q_\Phi\,X\right)^{1/3}-q_c\,X\right]\left[e^{1/b^2}\,\sqrt{\pi}\,{\rm Erf}\left(\frac{1}{b}\right)- 2\right]}
{2\,\gn\left[e^{n/b^2}\,q_c\,X+e^{1/b^2}\left(\frac{2\,n}{b^2}\left(1+6\,q_\Phi\,X\right)^{1/3}-q_c\,X\right)\right]^2}
\ ,
\ee 
where the dependence on the ADM mass is hidden in the compactness $X$.
This allows us to calculate the ratio
\be
\frac{M_0}{M}
=
\frac{3\,n\,e^{n/b^2}\left[\frac{2\,n}{b^2}\left(1+6\,q_\Phi\,X\right)^{1/3}-q_c\,X\right]\left[e^{1/b^2}\,\sqrt{\pi}\,{\rm Erf}\left(\frac{1}{b}\right)- 2\right]}
{2\left[e^{n/b^2}\,q_c\,X+e^{1/b^2}\left(\frac{2\,n}{b^2}\left(1+6\,q_\Phi\,X\right)^{1/3}-q_c\,X\right)\right]^2} \ .
\ee 
\par
Finally, we can rewrite the potential from Eq.~\eqref{VR_rho} as 
\be
V_{b}
&\!\!=\!\!&
\frac{n}{(n-1)\,q_c}
\ln\! \left[\frac{\frac{2\,n}{b^2}(1+6\,q_\Phi\,X)^{1/3}}{\frac{2\,n}{b^2}(1+6\,q_\Phi\,X)^{1/3}\!+q_c\,X\left(e^{(n-1)(1-r^2/R^2)/b^2}\!-\!1\right)}\right]
\nonumber
\\
&&
+
\frac{1}{4\,q_\Phi}
\left[
1-\left(1+6\,q_\Phi\,X\right)^{2/3}
\right]
\ ,
\label{VRfinal} 
\ee
where the suffix $b$ is to remark that this analytical expression stems form a density which only solves the polytropic
equation of state approximately due to $\rho_R>0$.
\par
Besides the compactness, this potential still depends on two parameters:
the width $b$ of the density distribution and the polytropic index $n$ from the equation of state.
Due to the complexity of the field equation~\eqref{EOMV}, we must rely on some approximate method
in order to study the dependence of the equation of state on the width of the density profile.
To this purpose, we write the potential as 
\be
V=V_b+ W
\ ,
\label{Vnr}
\ee
where $V$ is the exact solution to Eq.~\eqref{EOMV} and the difference $W$ with respect to the analytical
expression~\eqref{VRfinal} can be computed numerically.
The preferred values, or ranges of values, for the polytropic index $n$ will then be obtained by minimising the relative
error $W/V$ for given values of the width $b$ and compactness $X$.
In particular, we will perform the analysis for three values of the compactness:
small compactness $X=0.01$, intermediate compactness $X=0.1$ and large compactness $X=0.7$.
Considering the discussion of the discontinuity at the surface from the beginning of this Section, and the numerical
results from the previous Section, we are also interested in cases with $\rho_0\gg \rho_R$.
Therefore we will perform simulations for $b=0.5$, for which $\rho_R/\rho_0\simeq 1.8\%$.
In anticipation of the numerical results and plots, smaller values of $b$ result in much larger relative errors $W/V$
and do not represent good approximations.
For comparison, the relative error $W/V$ will also be shown for $b=1$, corresponding to a much larger
$\rho_R/\rho_0\simeq 37\%$.    
\begin{figure}[t]
\centering
\includegraphics[width=5.3cm]{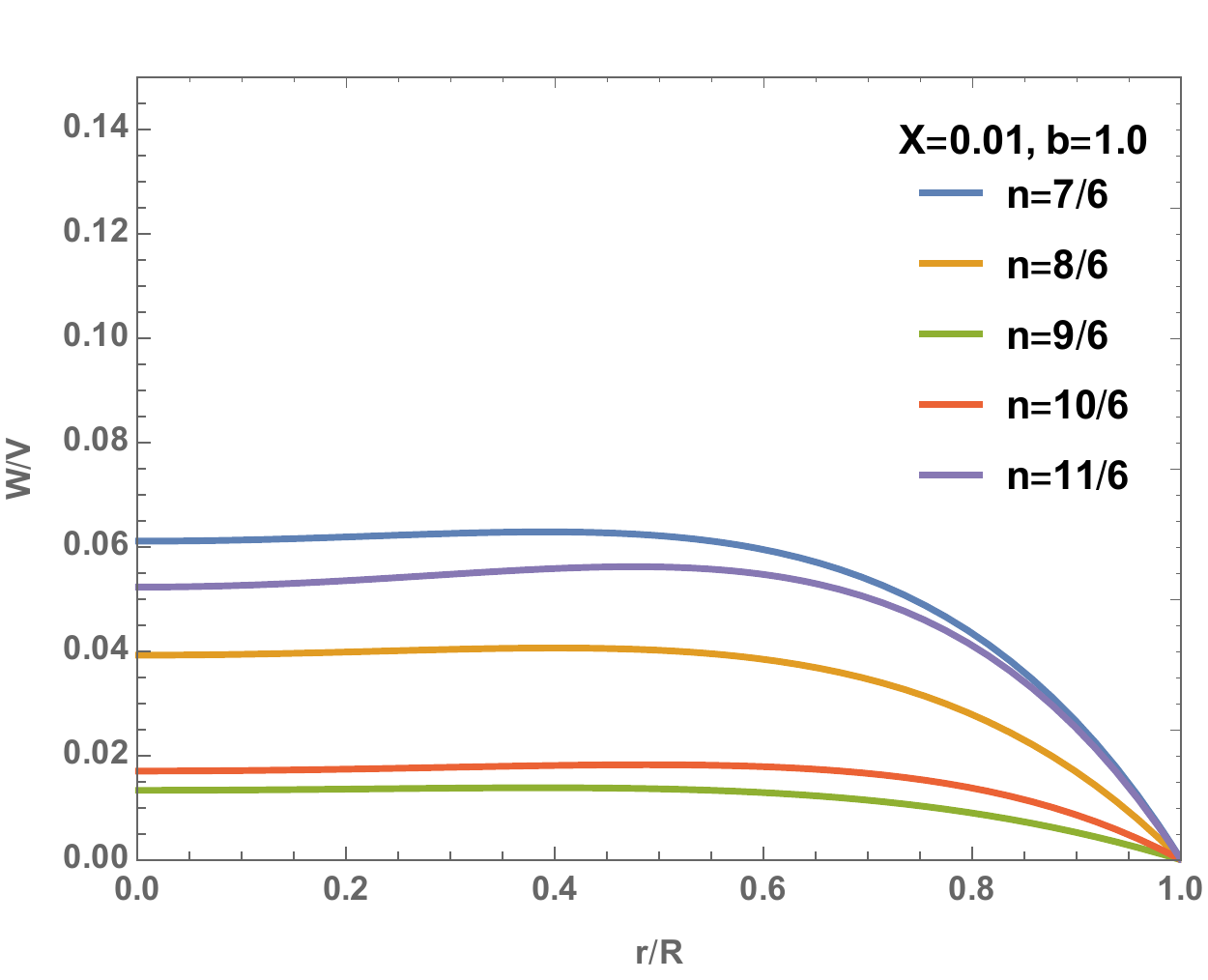}
\includegraphics[width=5.3cm]{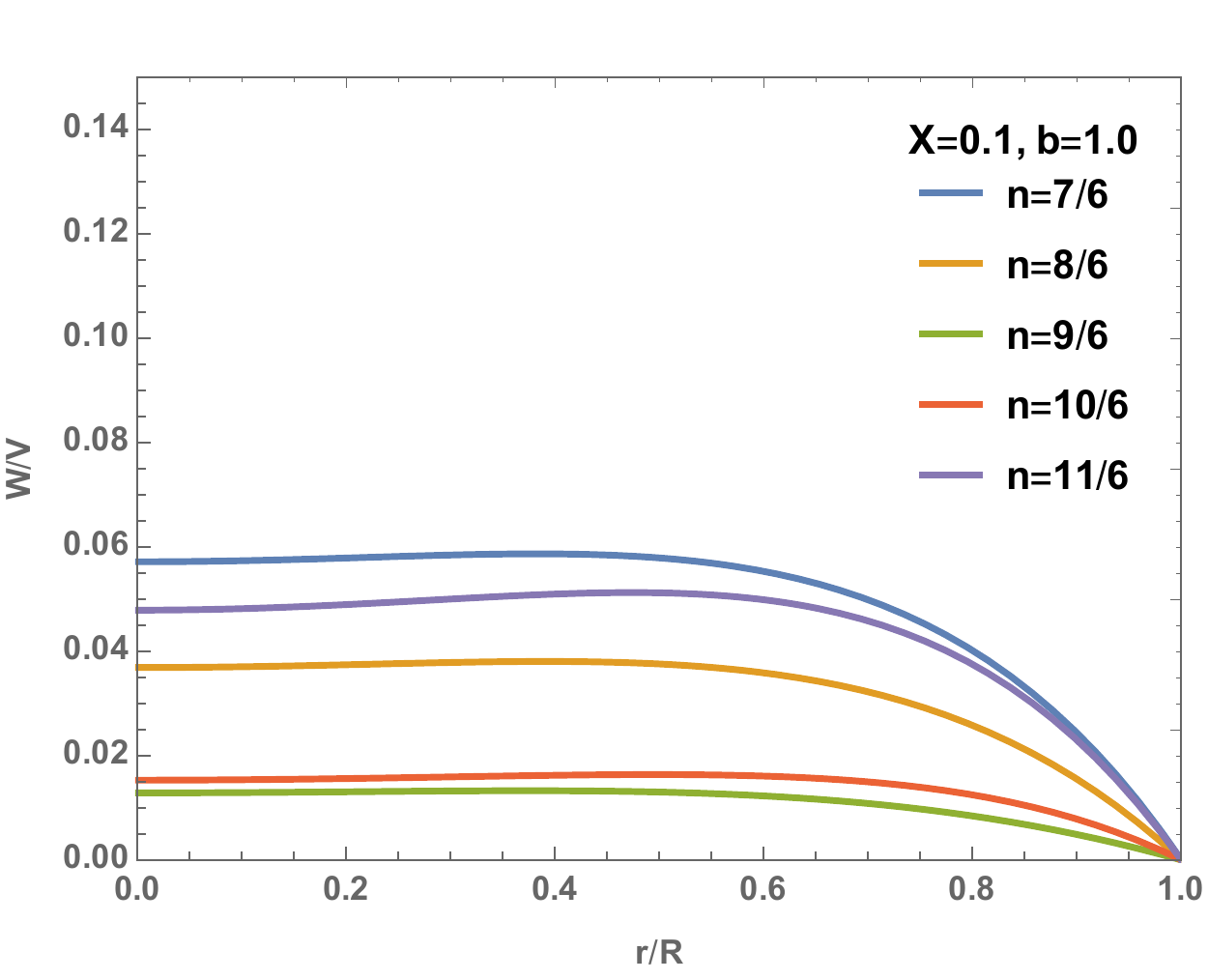}
\includegraphics[width=5.3cm]{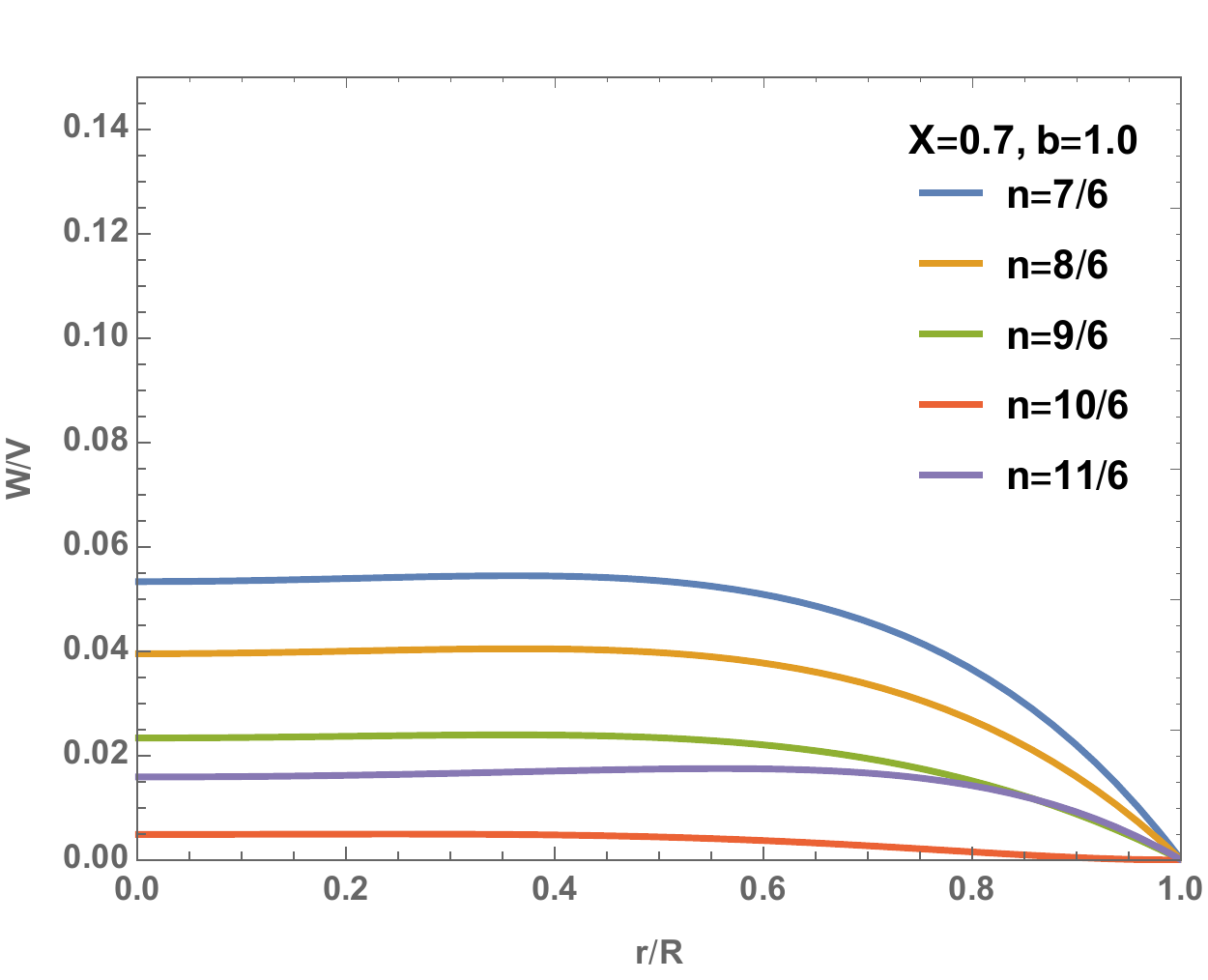}
\includegraphics[width=5.3cm]{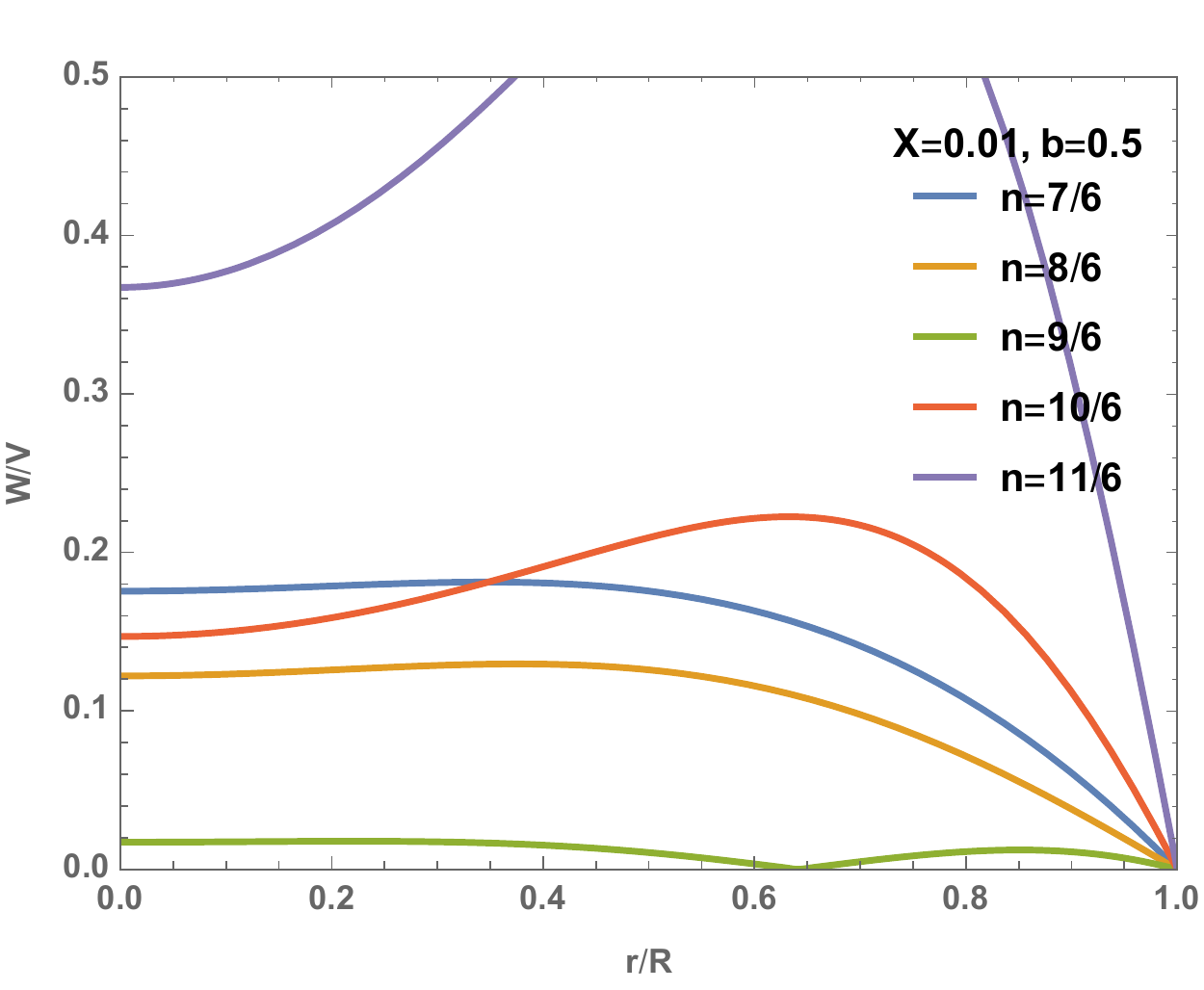}
\includegraphics[width=5.3cm]{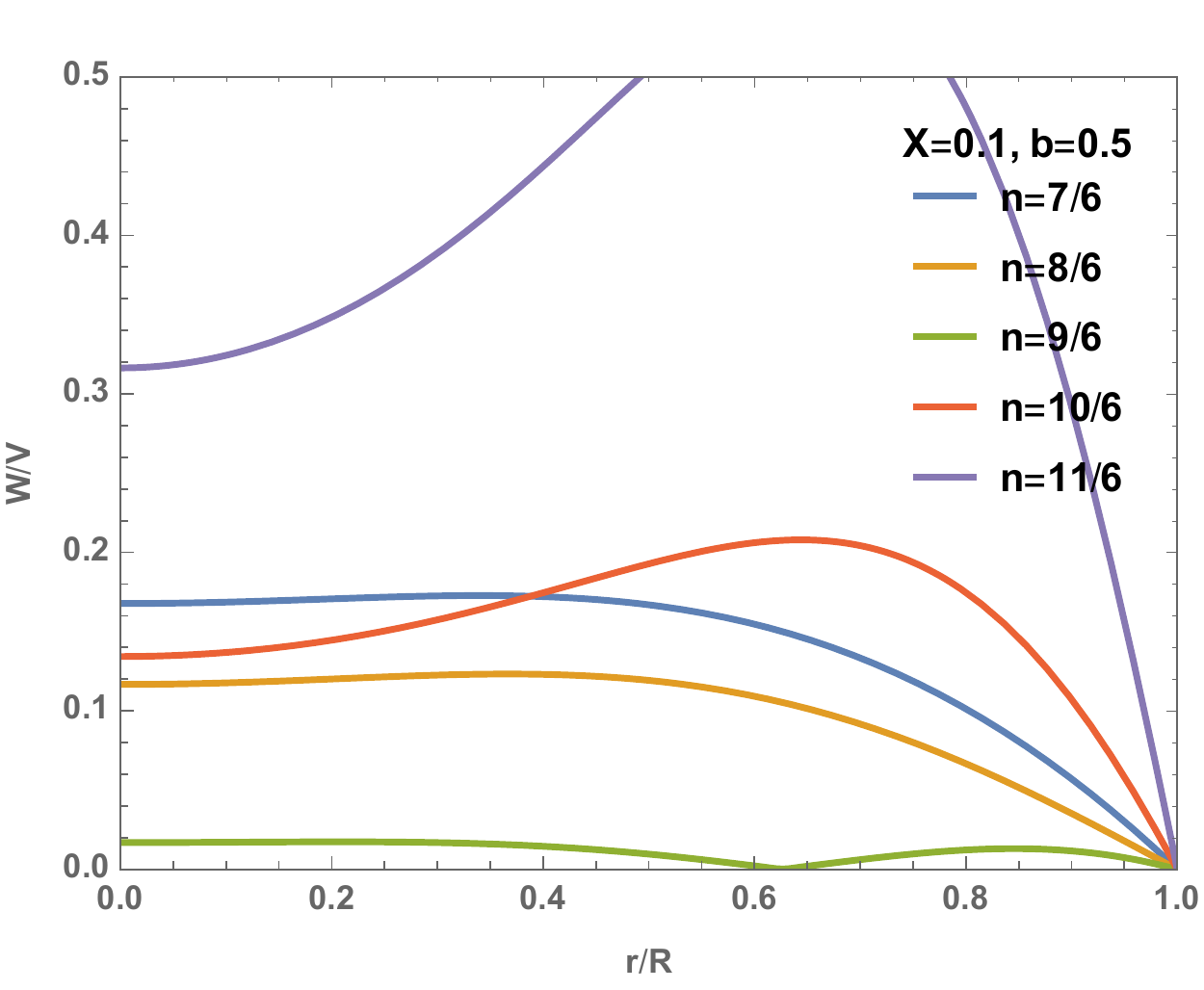}
\includegraphics[width=5.3cm]{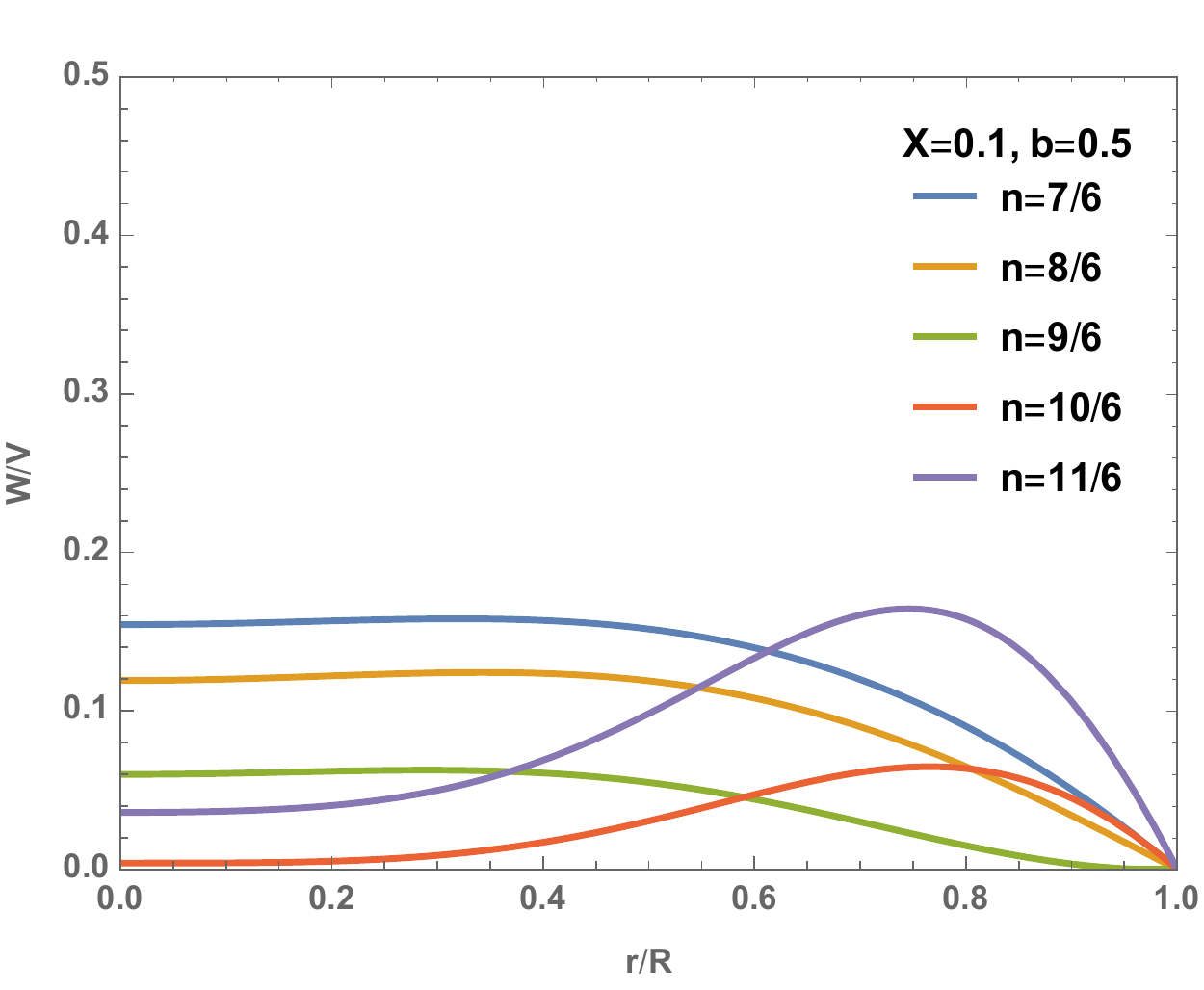}
\caption{Relative error for different values of the polytropic index $n$ for $q_c=q_\Phi=1$.
Note the different ranges on the vertical axis for the top plots versus the ones in the bottom.}
\label{error05}
\end{figure}
\par
In Fig.~\ref{error05} we display the relative errors for several values of the polytropic index covering the range
$1<n<2$ for all three values of the compactness and the two values of $b$ discussed earlier. 
In general, the relative error $W/V$ is smaller for larger values of $b$, corresponding to flat Gaussian profiles
with the density at the surface approximately equal to the one in the centre, thus departing
from our initial approximation. 
This general trend was also observed for other values of $b$ not included here. 
\par
With very few exceptions, the relative error grows to a maximum around the centre and vanishes at $r=R$. 
A simple explanation is that, while the parameter $\rho_0$ was obtained using only the leading order terms
in the series expansion of the field of motion around $r=0$, the boundary conditions at $r=R$ were matched
exactly. 
\par
For a fixed width $b=0.5$, the errors are the smallest for $n=3/2$, at least for small and intermediate compactness. 
In the high compactness case, even though the relative errors are always large, they become smaller for $n=5/3$,
which signals a possible transition from $n=3/2$ to $n=5/3$ as the compactness increases. 
\begin{figure}[t]
\centering
\includegraphics[width=10cm]{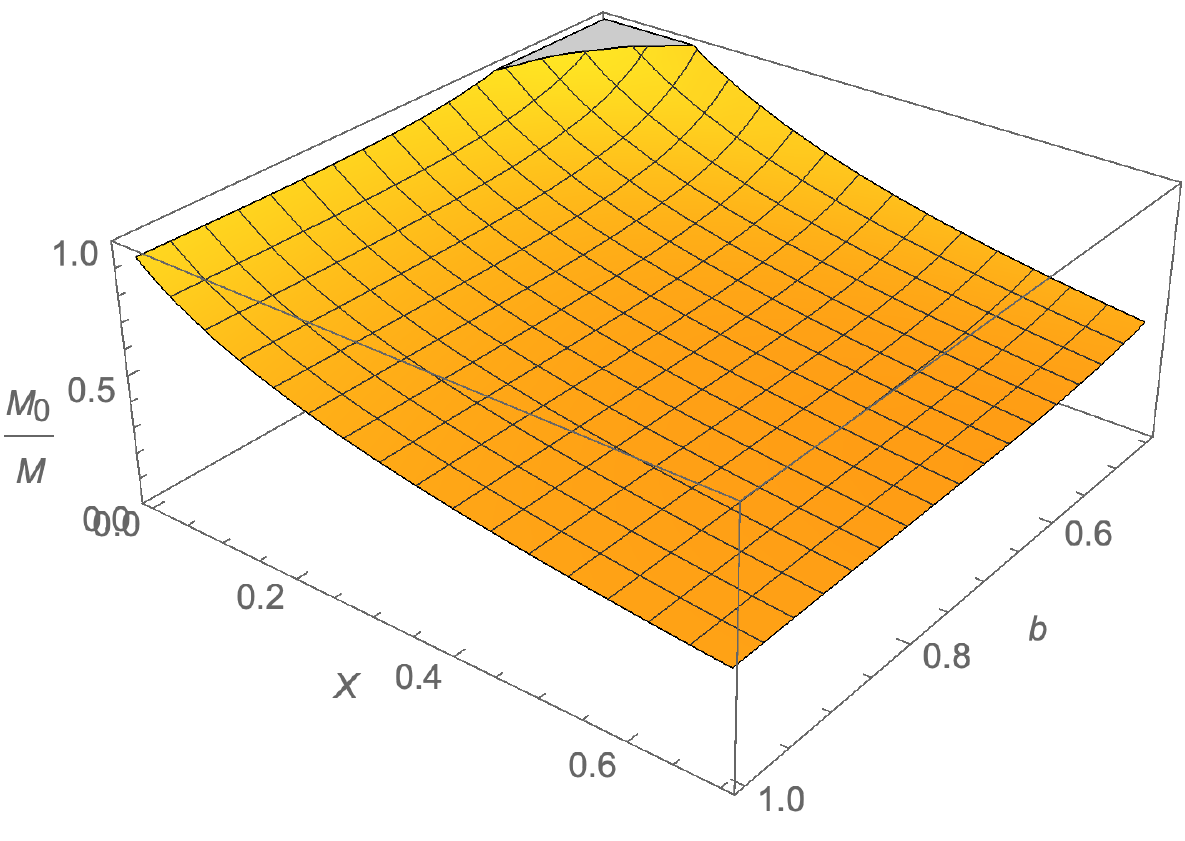}
\caption{Ratio ${M_0}/{M}$ as a function of the parameters $b$ and $X$ for $q_c=q_\Phi=1$ and $n=3/2$.}
\label{M0MbX}
\end{figure}
\par
The ratio $M_0/M$ for an equation of state with $n=3/2$ is presented in Fig.~\ref{M0MbX}.
Only values of $b>0.5$ are considered because, as stated before, the relative errors become unacceptably large
for smaller values of this parameter.
The limit of large $b$ is in agreement with our findings from Ref.~\cite{Casadio:2019pli}, where we investigated
objects with uniform densities. 
There it was found that the ratio $M_0/M<1$ for $q_\Phi=1$.
In fact, this ratio is smaller than one throughout most of the parameter space investigated here.
The only region in which the ratio becomes larger than one is for objects of low compactness with the density
peaked strongly near the center.  
In the low compactness limit, the ratio approaches one, as expected.
\par
\begin{figure}[t]
\centering
\includegraphics[width=5.3cm]{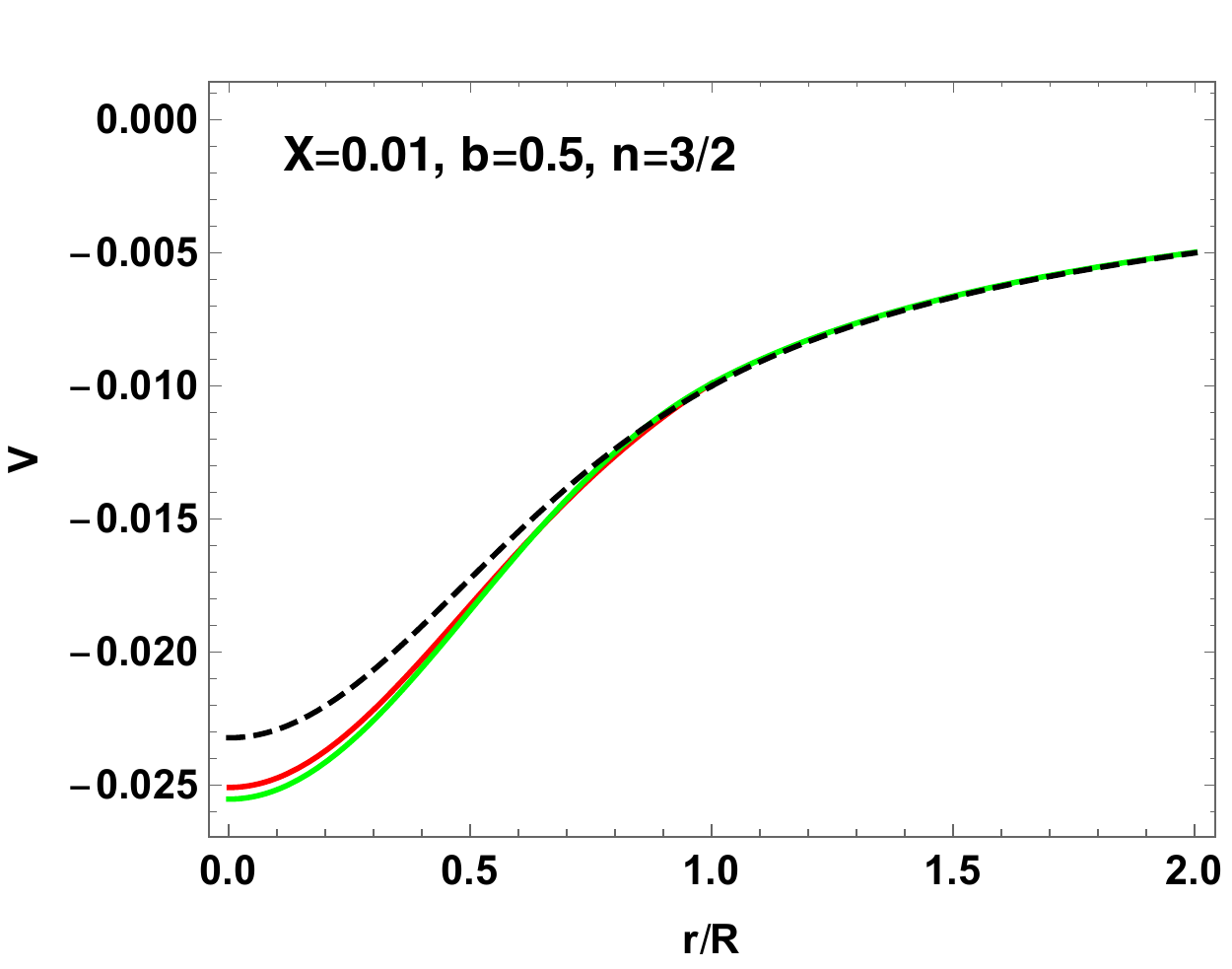}
\includegraphics[width=5.3cm]{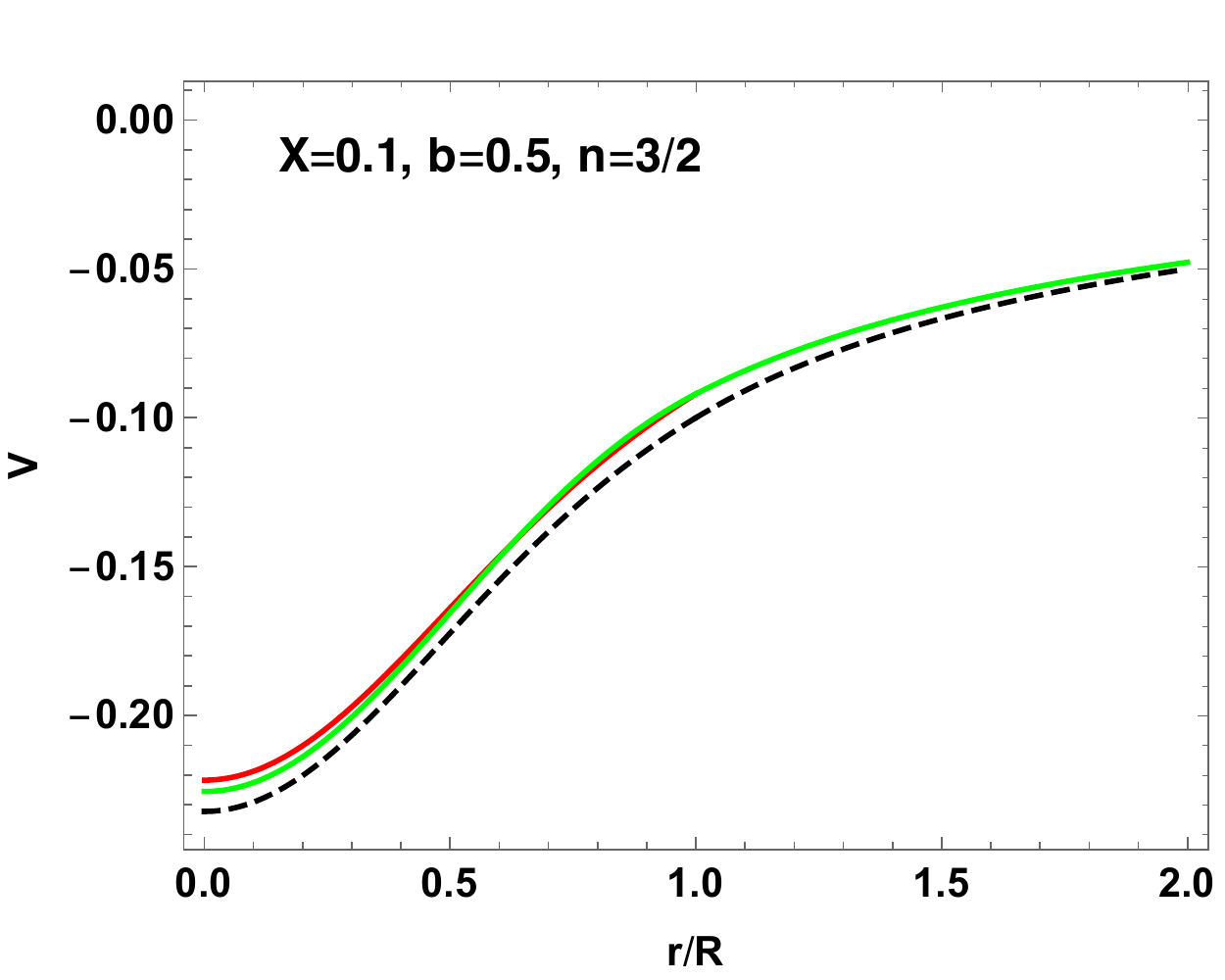}
\includegraphics[width=5.3cm]{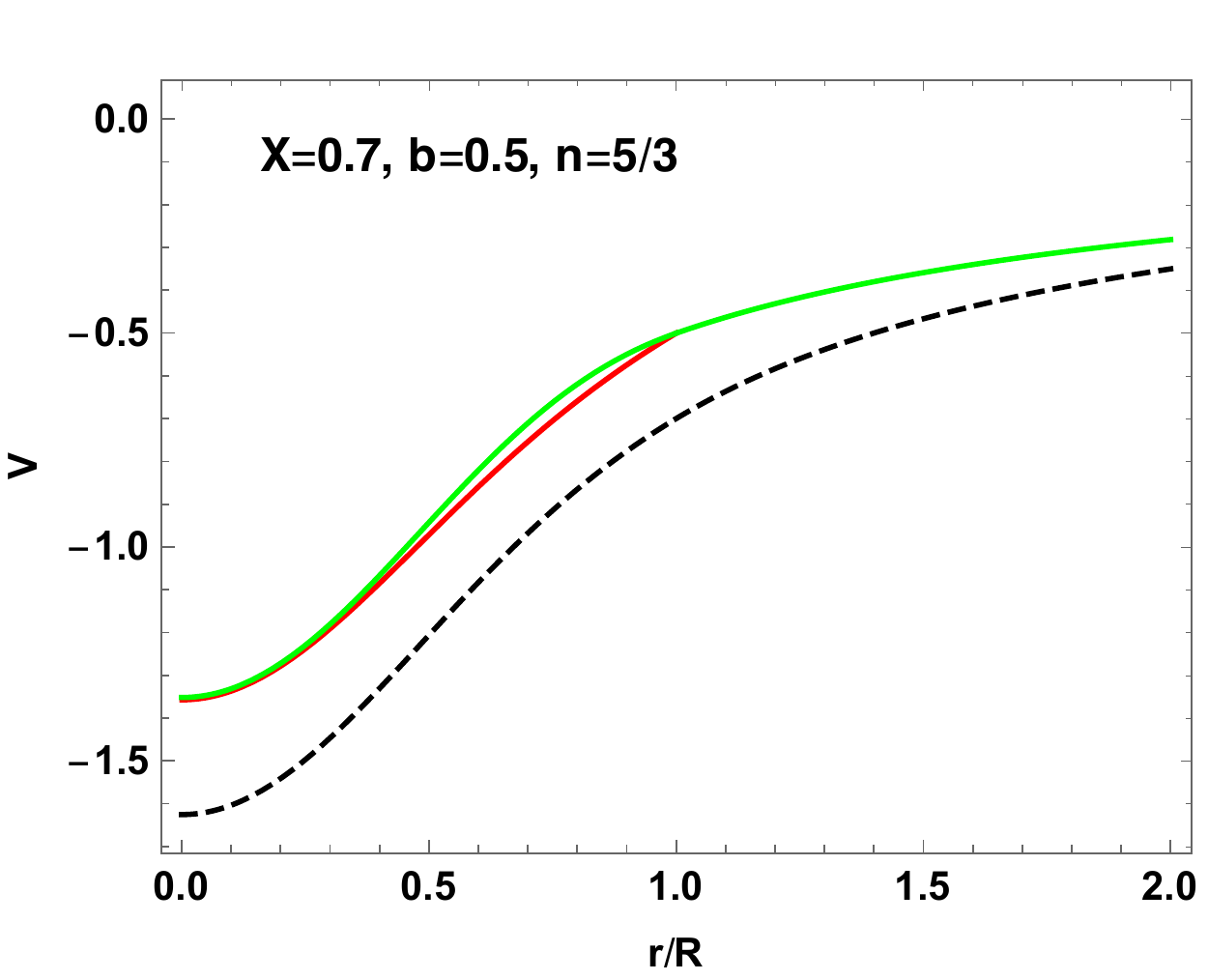}
\includegraphics[width=5.3cm]{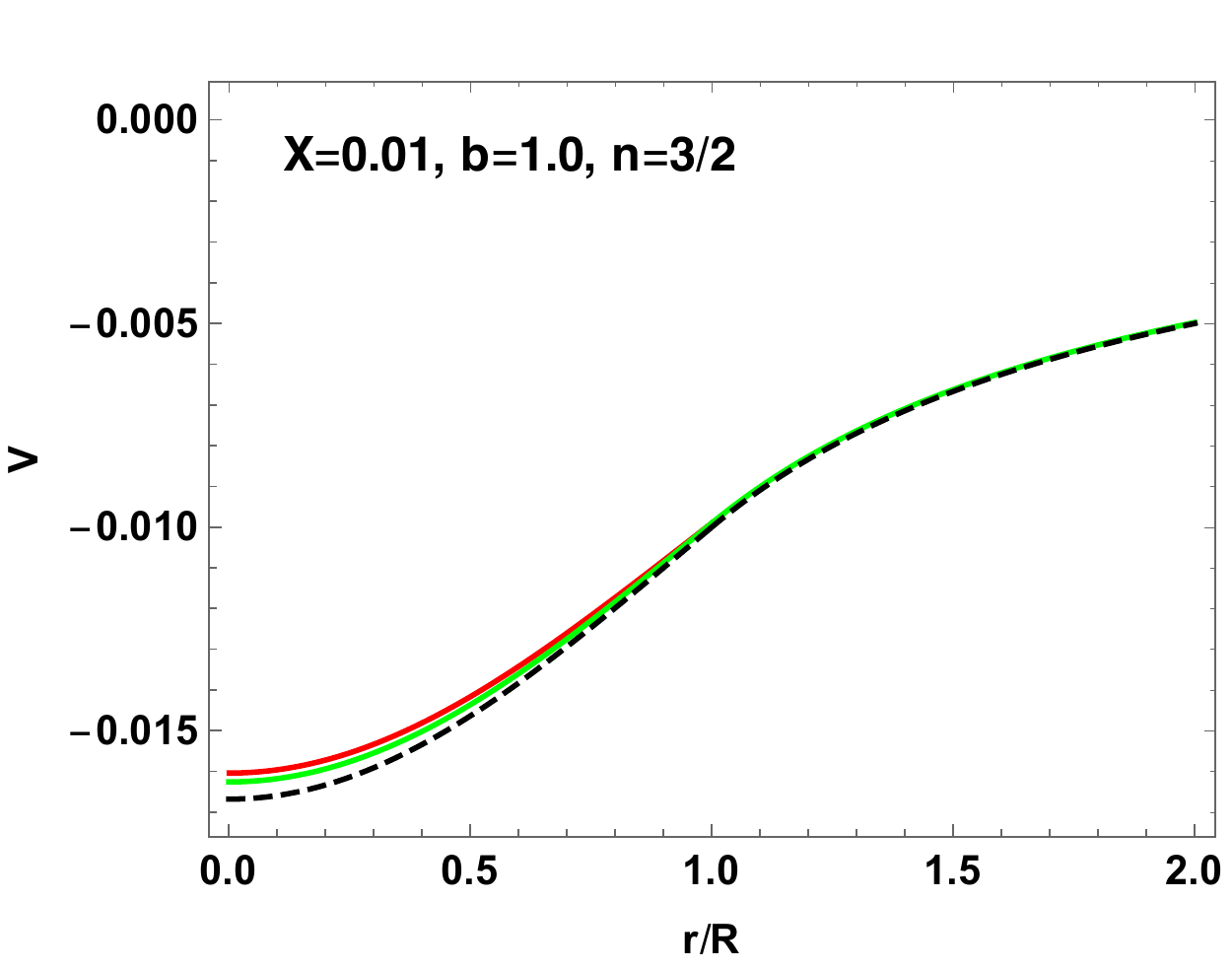}
\includegraphics[width=5.3cm]{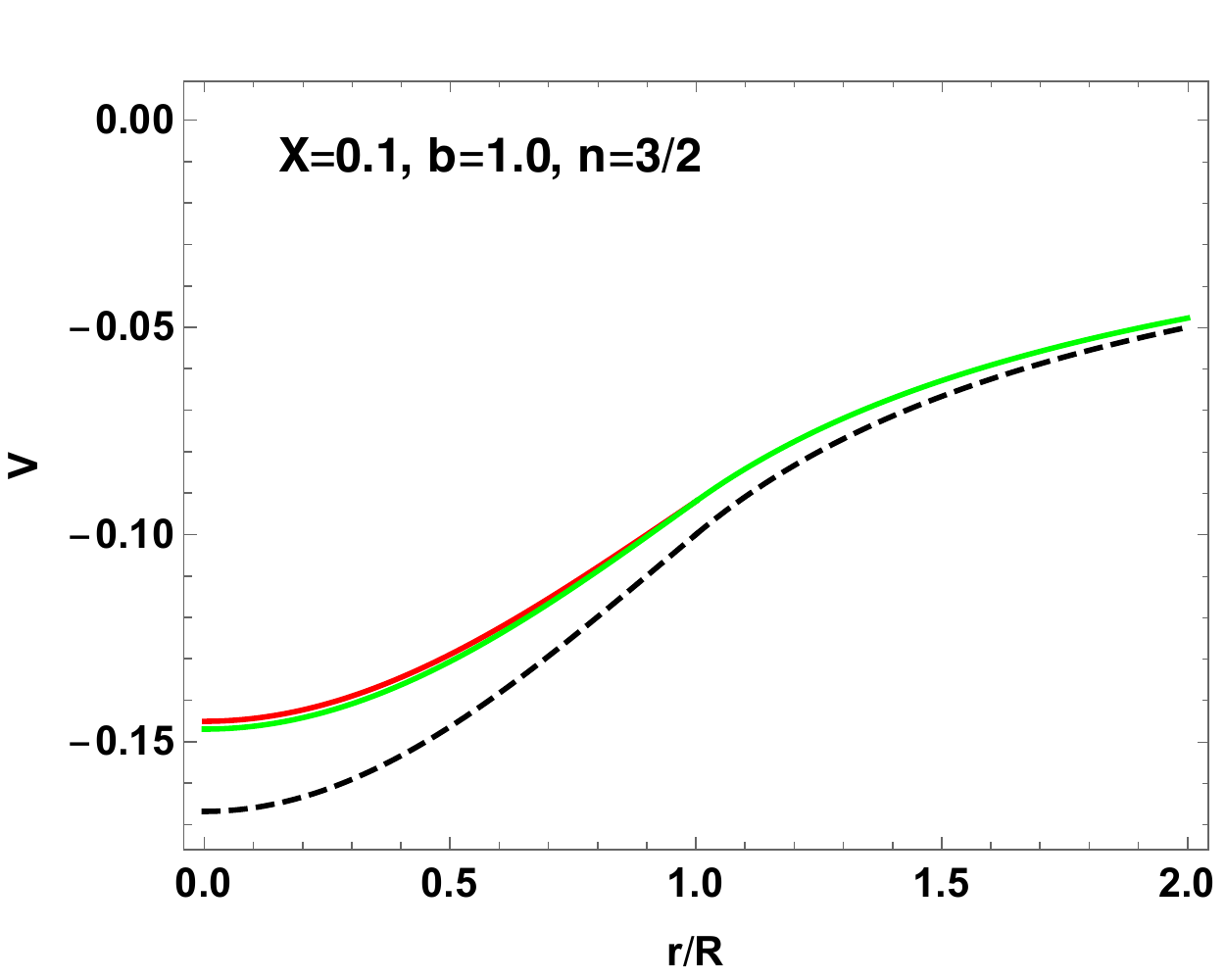}
\includegraphics[width=5.3cm]{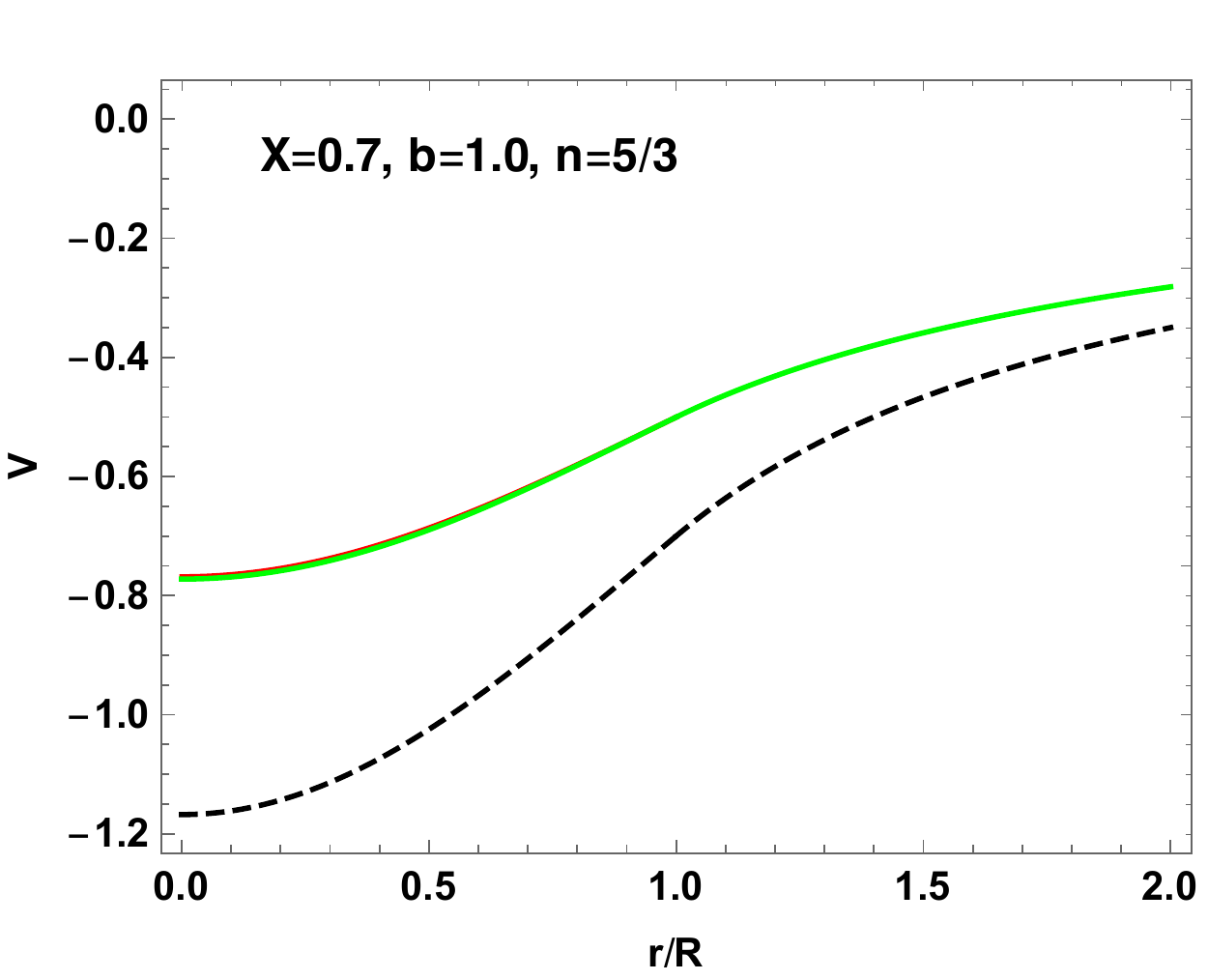}
\caption{Bootstrapped potentials for $q_c=q_\Phi=1$:
$V_b$ in Eq.~\eqref{VRfinal} (green lines) and $V$ in Eq.~\eqref{Vnr} (red lines).
For $X=0.01$ and $X=0.1$ we used $n=3/2$, while for $X=0.7$ we used $n=5/3$.
The dashed black lines represent the Newtonian potential $V_{\rm N}$ in Eq.~\eqref{VN_gaussian}
for a Gaussian matter distribution with the same $b$.}
\label{Vrplot}
\end{figure}

The gravitational potential from Eq.~\eqref{VRfinal} is plotted in Fig.~\ref{Vrplot} for the three values of $X$
considered here and for the values of $n$ which minimise the relative errors $W/V$.
Alongside, we also show the Newtonian potential corresponding to the same Gaussian distribution~\eqref{rho_gaussian},
given by
\be
V_{\rm N}
=
 \left\{
\begin{array}{ll}
\strut\displaystyle
-\frac{X\left[2\,r-\sqrt{\pi}\,b\,R\,e^{1/b^2}\,{\rm Erf}\left(\frac{r}{b\,R}\right)\right]}
{r\left[2-\sqrt{\pi}\,b\,e^{1/b^2}{\rm Erf}\left(\frac{1}{b}\right)\right]}
\ ,
&
r\leq R
\\
\\
\strut\displaystyle
-\frac{\gn\,M}{r} \,
\ ,
&
r>R
\ ,
\end{array}
\right.
\label{VN_gaussian}
\ee
for which we recall that the proper mass $M_0=M$.
The plots we obtain are consistent with our earlier findings.
The errors resulting from solving the equation of motion numerically are smaller for larger value of $b$,
represented in the three bottom plots, when compared to the corresponding plots obtained for the same
values of the compactness, but smaller $b$ values. 
As expected, the differences between the Newtonian and the bootstrapped Newtonian potentials
are larger for more compact objects and they becomes negligible as the density decreases.
The Newtonian potential generally creates deeper wells for most sources except for those characterised
by small values of the compactness and small values of the parameter $b$
(in our case for $X=0.01$ and $b=0.5$). 
Considering that $M_0=M$ in the Netwonian case, this signals that the ratio $M0/M$ crosses one,
as also observed in Fig.~\ref{M0MbX}.
However, we do not show here that, even in this case, we could find a value of the polytropic index $n$
slightly smaller than $n=3/2$, for which the ratio $M_0/M$ remains smaller than one, although
we cannot be certain that the same happens for general values of $X$ and $b$. 
\begin{figure}[h]
\centering
\includegraphics[width=7cm]{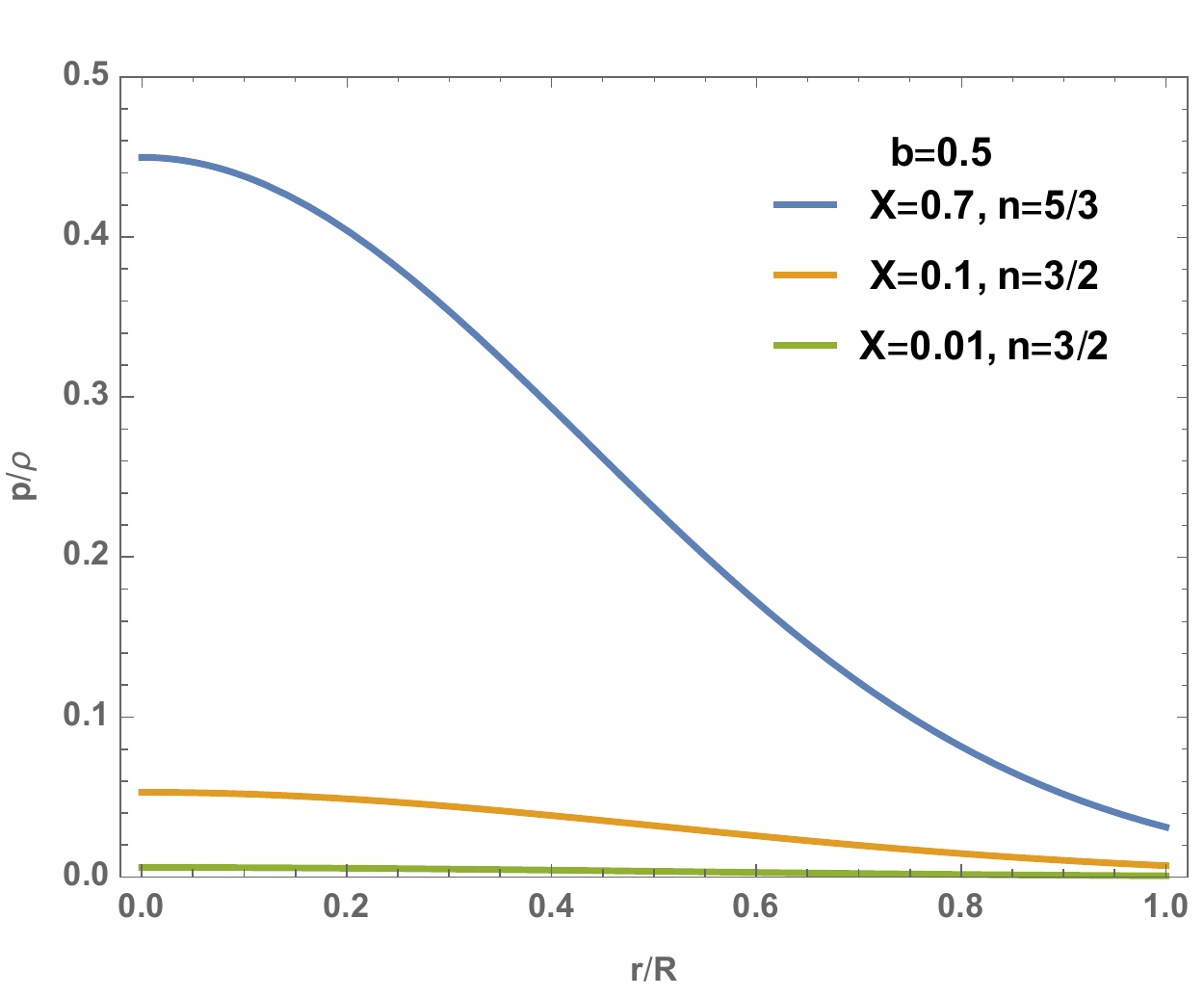}
$\quad$
\includegraphics[width=7cm]{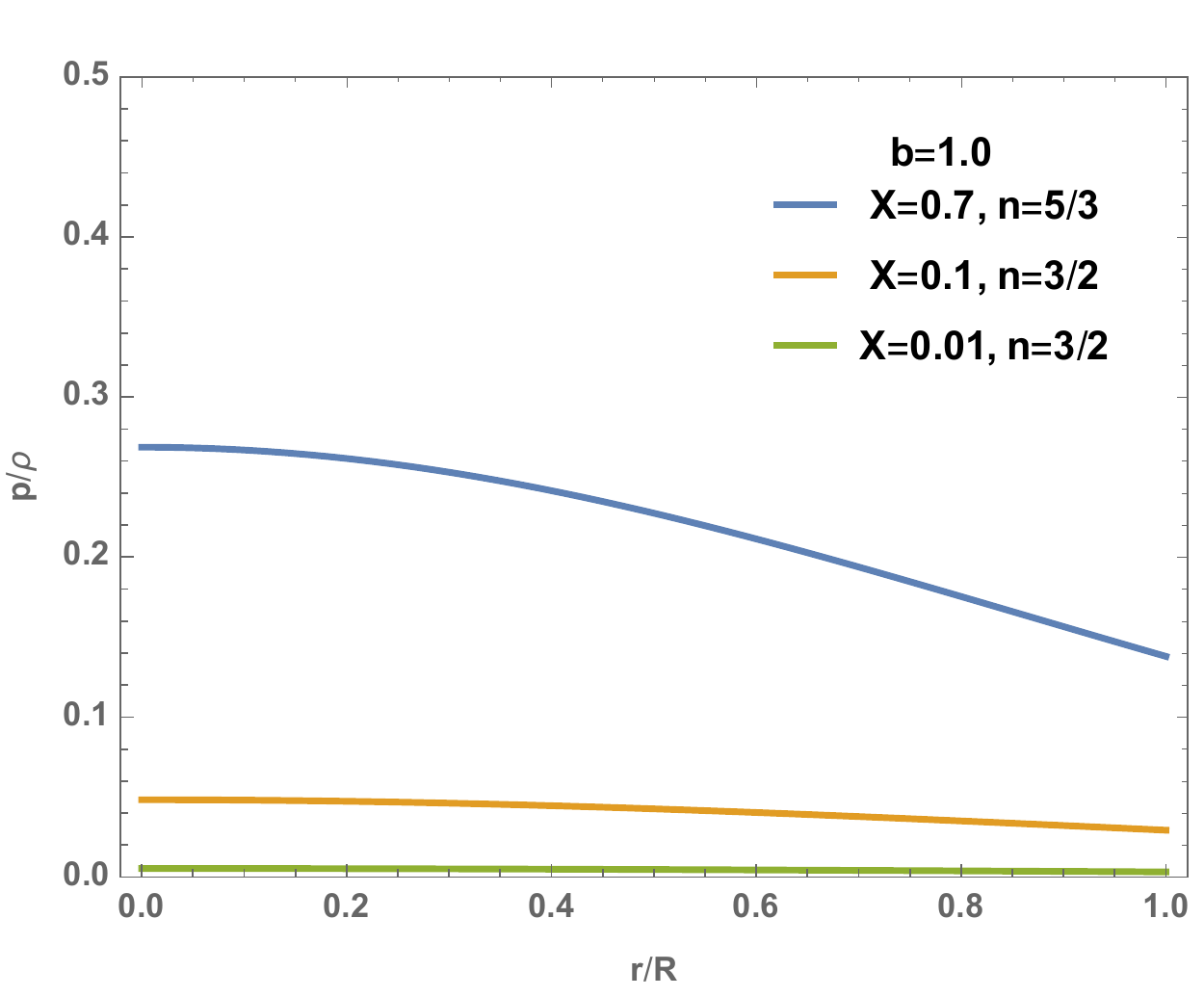}
\caption{Ratio of pressure to density for $q_c=q_\Phi=1$.}
\label{p_rho}
\end{figure}
\par
Finally, we show in Fig.~\ref{p_rho} the ratio
\be
\frac{p}{\rho}
=
\frac{X\,e^{(n-1)(1-r^2/R^2)/b^2}}{\frac{2\,n}{b^2}\left(1+6\,q_\Phi\,X\right)^{1/3}-q_c\,X}
\label{poverrho}
\ee
for each of the cases presented above.
For both values of $b$ taken into consideration, this ratio increases with the compactness.
What we observe is the expected behaviour in both limits:
while the pressure is negligible with respect to the density for objects of small compactness,
the ratio becomes larger as the objects become more compact, until the two quantities are
roughly of the same order of magnitude. 
\section{Comparison with General Relativity}
\label{S:tov}
\setcounter{equation}{0}
From the physical point of view, it is important to compare the solutions obtained in the bootstrapped Newtonian picture with
similar solutions of the Tolman-Oppenheimer-Volkoff (TOV) equation~\cite{tov} of General Relativity, namely
\be
r^2\,p'
=
-\gn\left(p+\rho\right)\left(m+4\,\pi\,r^3\,p\right)\left(1- \frac{2\,\gn\,m}{r}\right)^{-1}
\ ,
\label{eqTOV}
\ee 
where the Bondi mass function $m=m(r)$ is given by the same expression in Eq.~\eqref{proper_mass}. 
It is however important to notice that the radial coordinate $r$ in the bootstrapped Newtonian picture is 
associated with harmonic coordinates, and differs from the areal radius usually employed in the 
description of spherically symmetric systems in General Relativity.~\footnote{We use the same
symbol $r$ for both coordinates in order to keep the notation simpler, but a complete analysis of this
important issue requires deriving an effective bootstrapped metric which is currently being investigated~\cite{wip}.}
Correspondingly, the expression~\eqref{proper_mass}, which defines the proper mass 
in the bootstrapped Newtonian case, yields the Bondi mass of the star in General Relativity.
The latter, computed at the star surface of radius $r=R_{\rm TOV}$, equals the ADM mass
$m(R_{\rm TOV})=M_{\rm TOV}$ of the star~\cite{adm}.~\footnote{In this Section, we usually denote quantities
computed in General Relativity with the suffix TOV, in order to distinguish them from the analogous quantities
computed in the bootstrapped Newtonian picture.} 
\par
\begin{figure}[t]
\centering
\includegraphics[width=5.3cm]{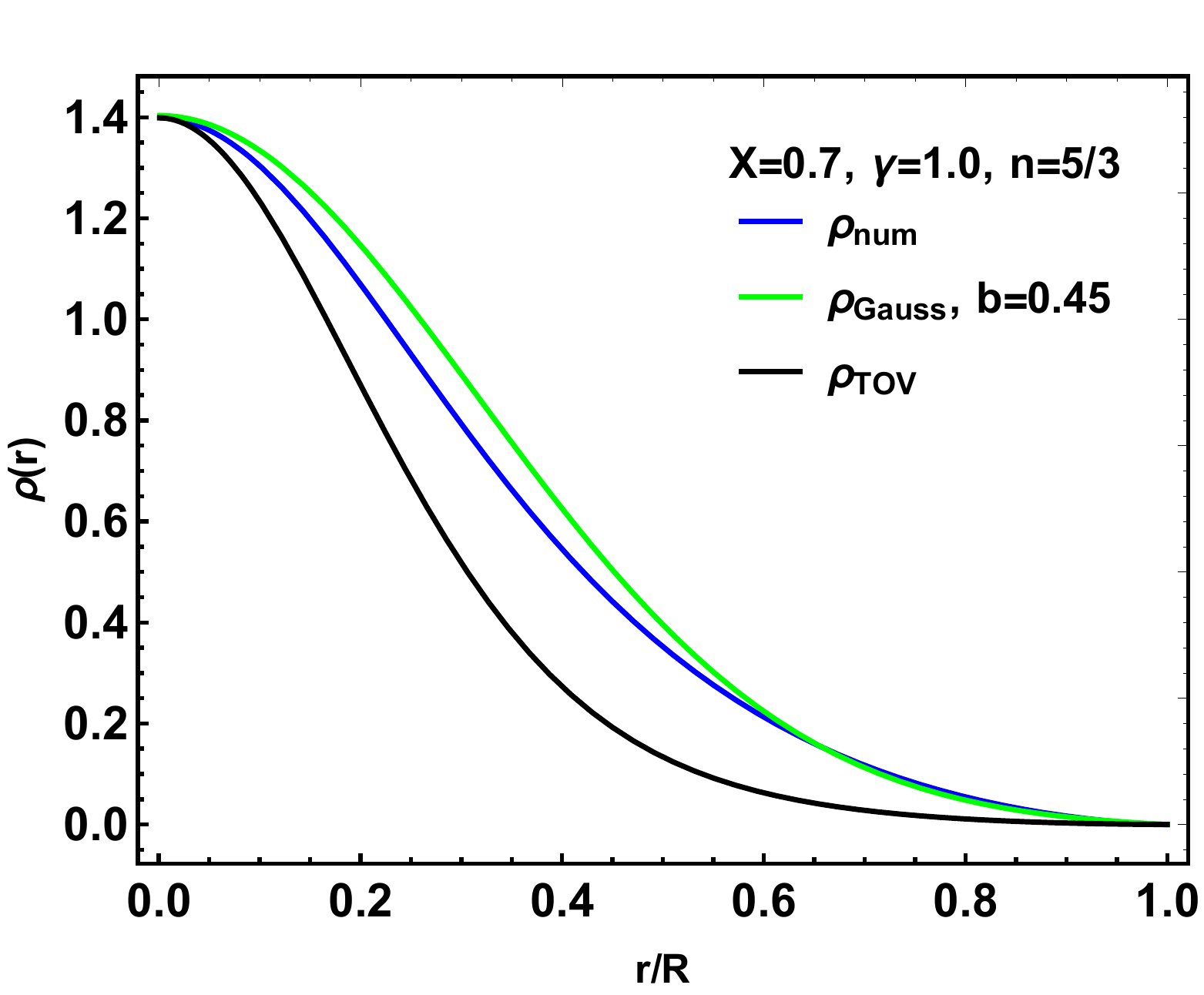}
\includegraphics[width=5.3cm]{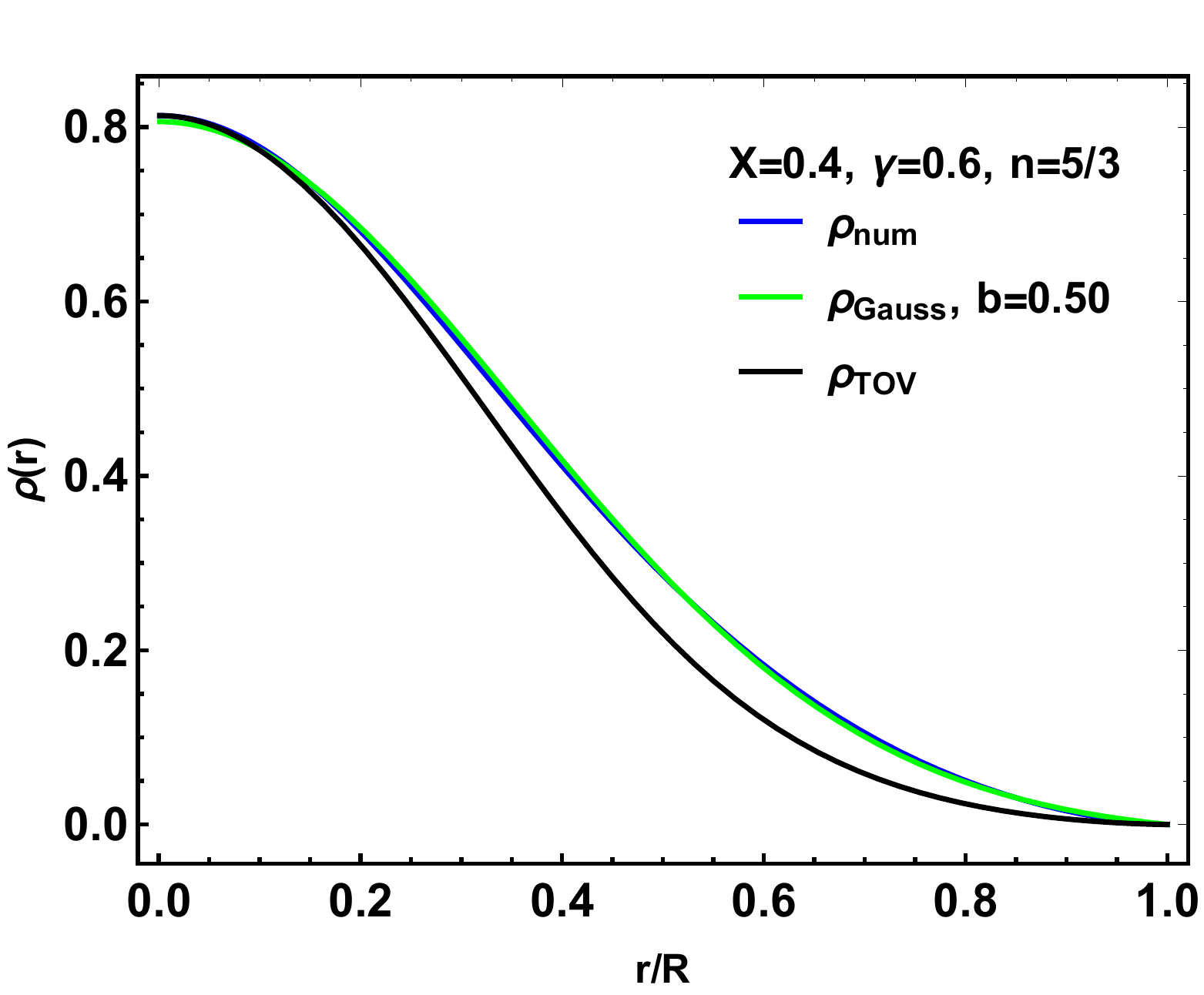}
\includegraphics[width=5.3cm]{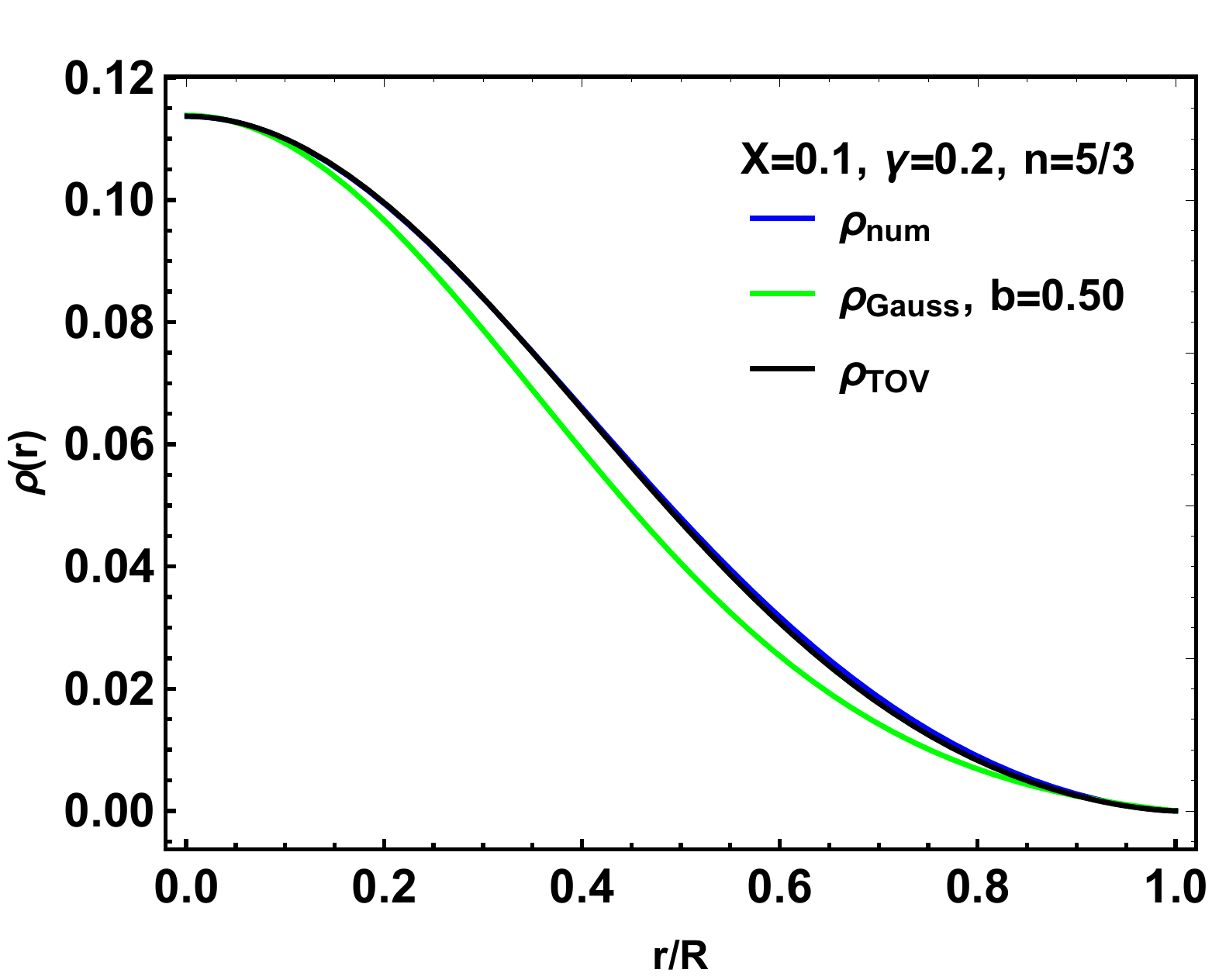}
\includegraphics[width=5.3cm]{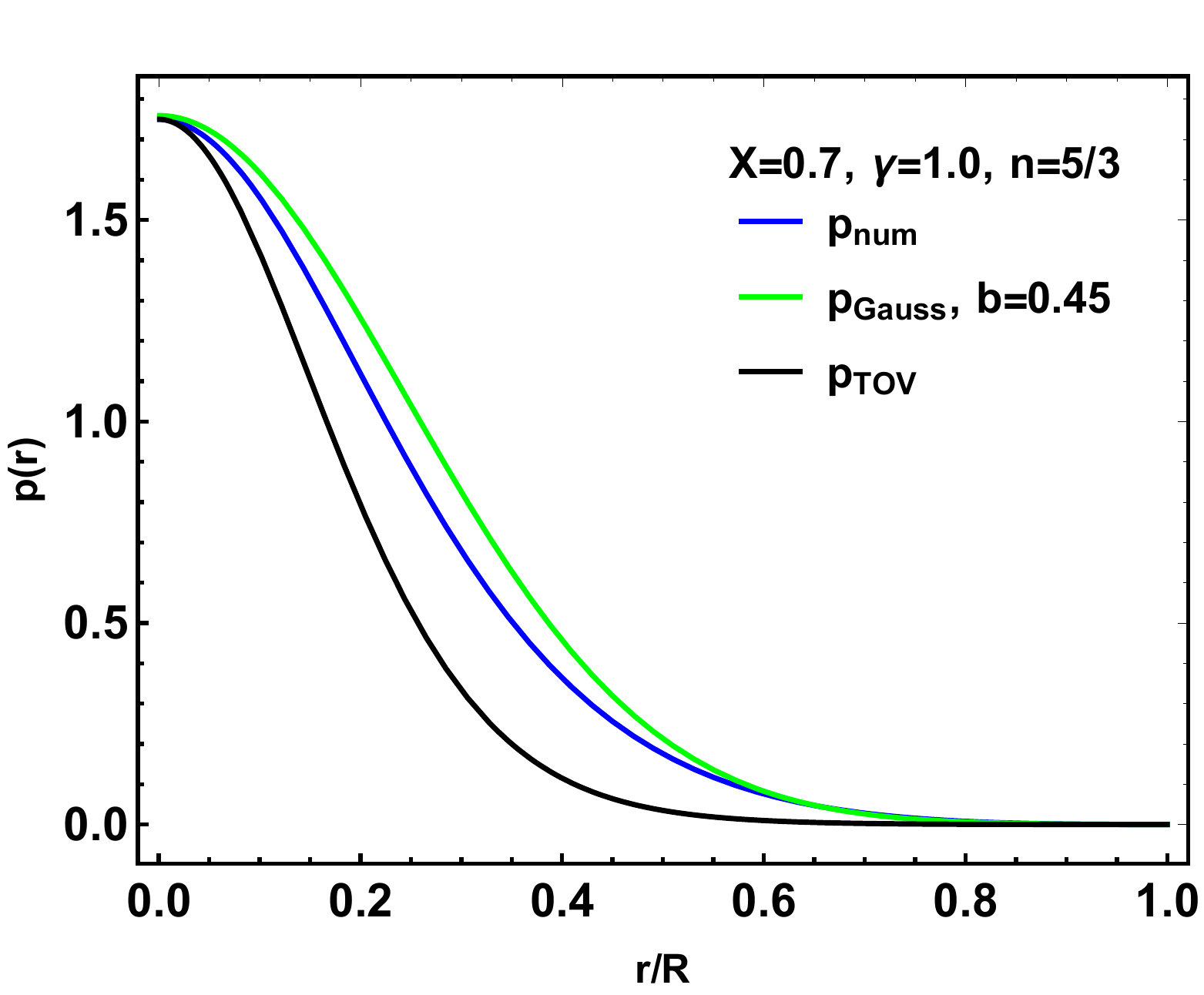}
\includegraphics[width=5.3cm]{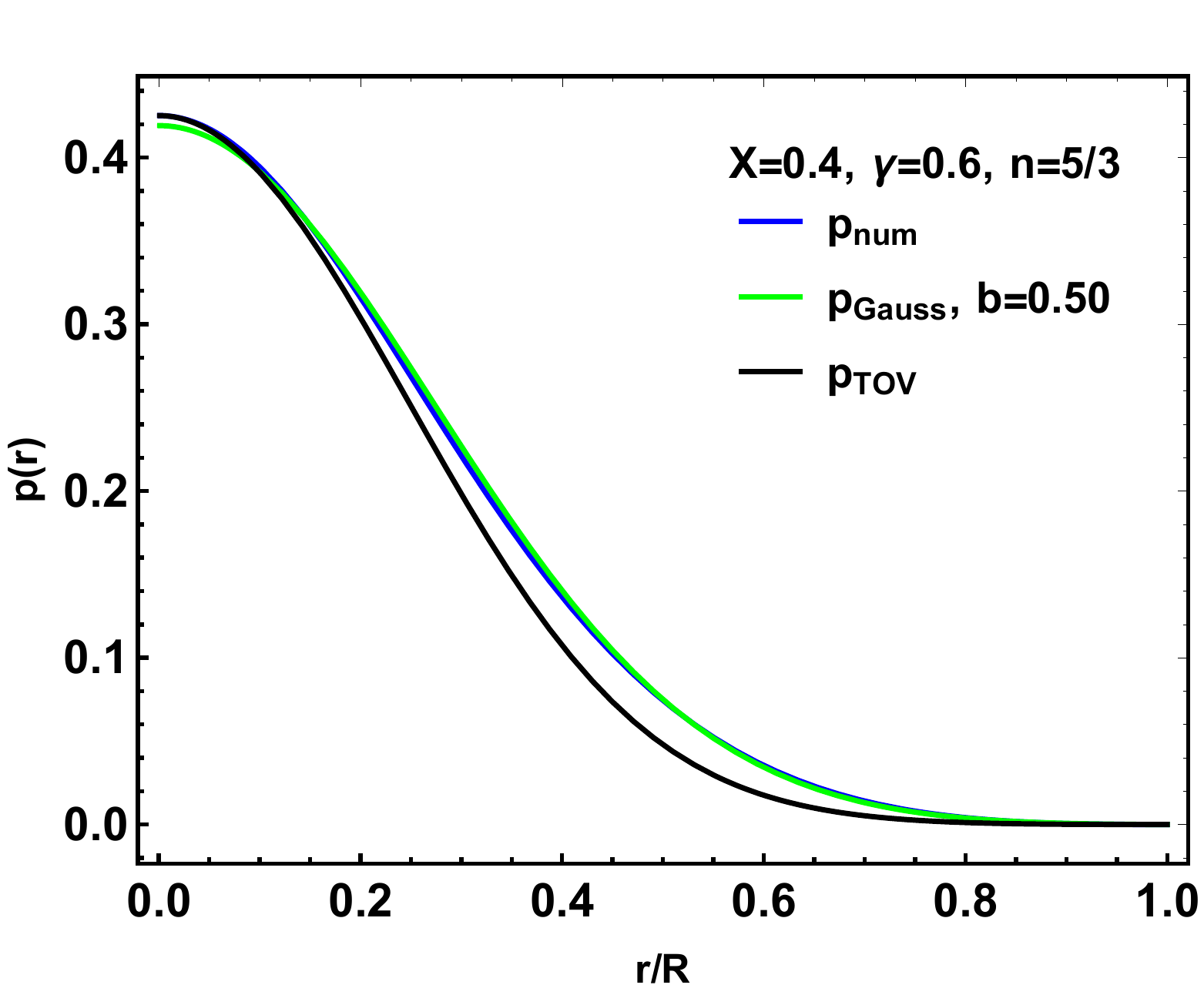}
\includegraphics[width=5.3cm]{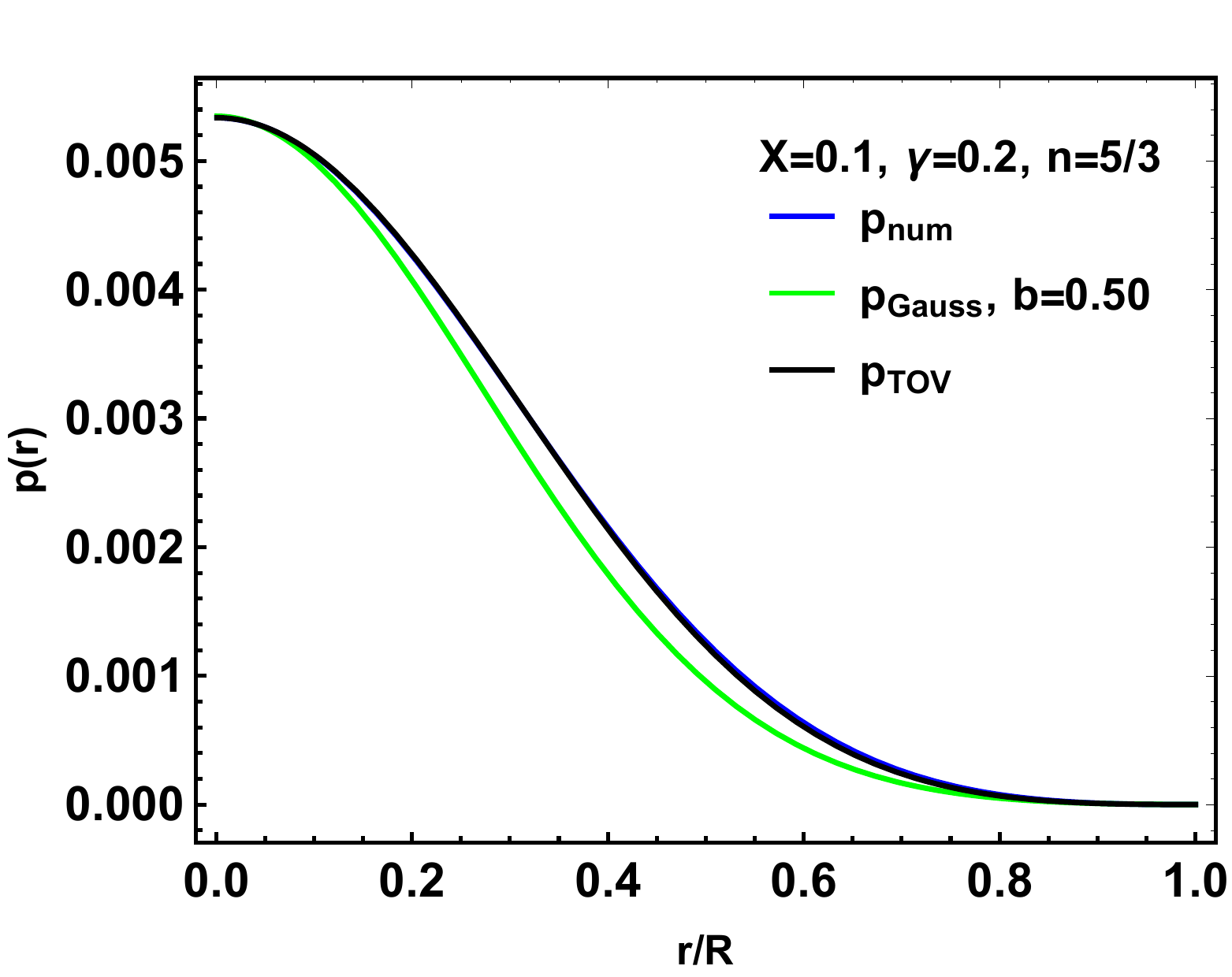}
\caption{Upper panels: density profiles obtained for polytropic stars by solving numerically the equation
for the density~\eqref{eomRho} in the bootstrapped picture (blue), the Gaussian density profiles
which approximate these solutions (green) and the  corresponding General Relativistic density profiles
(black).
Lower panels: pressure profiles for the same polytropic stars.
For all plots $q_c=q_\Phi=1$.}
\label{GR_comparison}
\end{figure}

Given a specific equation of state, the TOV equation~\eqref{eqTOV} determines the density profile $\rho=\rho(r)$
of the  compact source and can typically be solved only numerically.
In order to keep our comparison straightforward, we confront numerical solutions
for polytropic stars in the bootstrapped Newtonian picture, as they were obtained in Subsection~\ref{S:numerical},
with solutions obtained by solving numerically the TOV equation with the same equation
of state~\eqref{eosB} and central density $\rho_0=\rho(0)$. 
In particular, since the equation of state and $\rho_0$ are the same, the central pressure
$p_0=p(0)$ in the TOV solution also equals the central pressure in the bootstrapped Newtonian
solutions.
A few cases corresponding to different values of the compactness $X$ are shown in Fig.~\ref{GR_comparison},
where the pressure profiles $p=p(r)$ in the lower panels correspond to the density profiles $\rho=\rho(r)$
in the upper panels.
Also note that the bootstrapped Newtonian quantities are plotted as functions of $r/R$, with $R$
the corresponding star radius, whereas the TOV quantities are shown in terms of $r/R_{\rm TOV}$,
with $R_{\rm TOV}$ being the star radius obtained from solving Eq.~\eqref{eqTOV}.
We remark once more that $R$ and $R_{\rm TOV}$ in general differ, as we will discuss below.
For exemplification purposes, we also included the lines representing the Gaussian approximation.
The most relevant numerical quantities for the plotted cases are then displayed in Table~\ref{t:comparison}.
\par 
We first notice that we could find General Relativistic solutions throughout the entire bootstrapped
Newtonian compactness range $0.1\le X\le 0.7$, since the values of the corresponding TOV compactness 
are lower than the Buchdahl limit, $X_{\rm TOV}< X_{\rm BL}=4/9$.
The General Relativistic density profiles are always below the curves obtained numerically
in the bootstrapped picture (we will refrain from comparisons with the Gaussian approximation
since in that case a certain degree of arbitrariness exists in the choice of the parameter $b$).
This is consistent with the ratio of the masses $M_{\rm TOV}/M$ always being smaller than one.
As the compactness $X$ decreases, this ratio becomes closer to one and the two density profiles
become virtually identical. 
Moreover, the radius $R$ of a bootstrapped Newtonian star is also usually larger than the
$R_{\rm TOV}$ of the General Relativistic polytrope. 
The fact that $X>X_{\rm TOV}$ means that the bootstrapped picture can allow for the existence of more
compact (either smaller in size or more massive) polytropic stars than General Relativity.
When looking at the lower panels of Fig.~\ref{GR_comparison} in comparison to the upper ones,
we see that  just like in the case of the density profiles, the pressure inside bootstrapped Newtonian
polytropes is always larger (for high compactness values) or at least equal (for low compactness)
to the pressure inside the corresponding General Relativistic cases. 
\begin{table}[h]
\centering
\begin{tabular}{||c|c|c|c|c||c|c|c|c|||}
\hline\hline
$X$ & $n$ & $\bar\gamma$ & $\gn\,\rho_0\,R^2$ & $\gn\,p_0\,R^2$ & $R_{\rm TOV}/R$  & $M_{\rm TOV}/M$ & $X_{\rm TOV}$ & $X_{\rm TOV}/X$ \\ \hline
\hline
$0.7$ & $5/3$  & $1.0$ & $1.40$ & $1.75$ & $0.92$ & $0.35$ & $0.23$ & $0.33$ \\ 
\hline
0$.4$ & $5/3$ & $0.6$ & $0.81$ & $0.42$ & $0.81$ & $0.39$ & $0.23$ & $0.58$ \\ 
\hline
$0.1$ & $5/3$ & $0.2$ & $0.11$ & $0.005$ & $0.95 $ & $ 0.83$ & $0.070$ & $ 0.70$ \\ 
\hline
\hline
\end{tabular}
\caption{Comparison between bootstrapped Newtonian parameters and TOV solution for the cases in Fig.~\ref{GR_comparison}.}
\label{t:comparison}
\end{table}
\section{Conclusions and outlook}
\label{S:conc}
\setcounter{equation}{0}
We have extended our model of bootstrapped Newtonian stars from Ref.~\cite{Casadio:2019cux}
by investigating objects with non-uniform mass distribution.
We assumed a polytropic equation of state to determine the relation between the density and the pressure
inside the star.
In order to avoid singular configurations, we imposed the condition for the first derivative of the density
to vanish in the centre of the star.
The other boundary condition was for the density to vanish at the surface.
Starting from the equation of continuity and the polytropic equation, we could rewrite the field equation
in terms of the density and its derivatives. 
This was then solved numerically to obtain the density profile for various values of the compactness
$X$ of the star, the parameters $\gamma$ and $n$ which define the polytropic state and the coupling
constants in our model (which we usually set to one for simplicity). 
Ideally, in order to draw general conclusions, one should scan the entire parameter space, which 
is a very demanding numerical task we prefer to leave for future developments.
However, the numerical results presented in Section~\ref{S:numerical} show density profiles which can be
approximated by Gaussian distributions fairly accurately, therefore, we focused on these solutions next. 
\par
Starting from a Gaussian density profile, along with the polytropic equation of state, one can describe
the star in terms of the width $b$ of the distribution, the polytropic index $n$ and the compactness $X$
of the object.
Analytic approximations can then be obtained for the gravitational potential.
The accuracy of these analytic approximations was then estimated numerically in order to determine
the polytropic index $n$ compatible with such distributions.
It appears that larger values of $n$ are favoured for larger compactness.
We also compared the proper mass $M_0$ to the ADM mass $M$, and found that $M_0<M$
throughout most of the parameter space.
We recover our previous results obtained for uniform density profiles in the large $b$ limit.
There is however a small region of the parameter space where, at least for some values of the
polytropic index $n$, the proper mass $M_0>M$.
In this regime, we also found that the bootstrapped potential creates a deeper potential well than
the Newtonian one, while in all other cases the opposite occurs (see Fig.~\ref{Vrplot}). 
The total gravitational potential energy is computed in Appendix~\ref{S:energy} and
turns out to be ngative for all solutions analysed.
Finally, pressure and density were found to have the expected behaviour.
In particular, their ratio increases with compactness $X$ and with the polytropic index $n$,
the former having a greater impact.
\par
We also compared the density and pressure profiles for bootstrapped Newtonian,
respectively General Relativistic, polytropes with the same equation of state~\eqref{eosB}
and central density $\rho_0=\rho(0)$.
We emphasise once more that, while the Newtonian limit is obtained when the couplings 
$q_\Phi$ and $q_c$ in Eq.~\eqref{EOMV} are equal to zero,
the bootstrapped Newtonian gravity is expected to yield the results closest to those of
General Relativity for $q_\Phi\simeq q_c\simeq 1$, which is the case investigated here.
The preliminary results obtained for these types of stars show that the bootstrapped Newtonian
gravity allows for objects with larger densities (and therefore masses).
At least in the relatively high compactness regime considered in Section~\ref{S:tov}, the bootstrapped picture leads to 
more compact stars than those obtained by solving the TOV equation.
In other words, the bootstrapped Newtonian gravity makes a significant difference in the
high compactness regime because it can accommodate for the existence of more compact and massive
self-gravitating stars than General Relativity, given the same polytropic equation of state
and central density.
Of course, these more compact and massive stars are also balanced by larger pressure
values.
In the opposite limit of low compactness, the density profiles predicted in the bootstrapped 
Newtonian picture become practically identical to those of General Relativity.
The phenomenological differences between the two theories therefore manifest only for highly
compact objects, consistently with the original motivation of accounting for the quantum 
nature of matter and gravity at large compactness, a task hard to tackle starting from full General Relativity.
A more in depth comparison of the bootstrapped Newtonian and General relativistic stars will be the
subject of future work.
The comparison will be extended to not only polytropes, but also to objects governed by other equations of state.
\par
We would like to conclude by recalling that the bootstrapped description for compact self-gravitating
objects was mainly developed with the purpose of investigating the corpuscular description of quantum
gravity originally put forward for black holes~\cite{DvaliGomez,Giusti:2019wdx,Dvali:2016mur,kuhnelBaryons}.
In fact, the absence of a Buchdahl limit allows for (Newtonian) horizons surrounding a matter core of large
but finite compactness.  
it was then shown in Ref.~\cite{Casadio:2020mch} that the bootstrapped potential for uniform
sources admits a description in terms of a coherent quantum state of gravitons, provided the matter
source is also described by quantum physics.
It will therefore be interesting to widen the survey of the parameter space for polytropic stars
here initiated in light of more explicit quantum descriptions of matter.
\section*{Acknowledgments}
R.C.~is partially supported by the INFN grant FLAG and his work has also been carried out in
the framework of activities of the National Group of Mathematical Physics (GNFM, INdAM)
and COST action {\em Cantata\/}. 
O.M.~is supported by the grant Laplas~VI of the Romanian National Authority for Scientific
Research. 
\appendix
\section{Gravitational energy}
\label{S:energy}
\setcounter{equation}{0}
The gravitational potential energy $U_{\rm G}$ in the bootstrapped picture can be estimated from the effective
Hamiltonian given in Eq.~\eqref{HamV}, which we separate into three contributions as~\cite{Casadio:2019cux} 
\be
U_{\rm G}= U_{\rm BG}+U_{\rm GG}^{\rm in} +U_{\rm GG}^{\rm out}
\ ,
\label{UG}
\ee
where
\be 
\label{UBG}
U_{\rm BG}
&\!\!=\!\!&
4\,\pi
\int_0^\infty
r^2\,\d r
\left(\rho+q_c\,p\right) V\left(1-2\,q_\Phi\,V\right) \ ,
\\ \label{UGGin}
U_{\rm GG}^{\rm in} 
&\!\!=\!\!&
\frac{1}{2\,\gn}
\int_0^R
r^2\,\d r\,
\left(V_{\rm in}'\right)^2
\left(1-4\,q_\Phi\,V_{\rm in}\right)  \ ,
\\ \label{UGGout}
U_{\rm GG}^{\rm out} 
&\!\!=\!\!&
\frac{1}{2\,\gn}
\int_R^\infty
r^2\,\d r\,
\left(V_{\rm out}'\right)^2
\left(1-4\, q_\Phi\,V_{\rm out}\right)
\ .
\ee
The contribution from the outer vacuum is exactly given by
\be
U_{\rm GG}^{\rm out} 
= 
\frac{\gn\,M^2}{2\,R}
\ , 
\ee
while the inner contributions can only be evaluated numerically within the approximations
for the potential employed in the previous sections.
\par
Siince the potential $V$ is negative and has positive slope everywhere,
one can see from their expressions above that the ``baryon-graviton'' component $U_{\rm BG}$ is negative,
whereas the ``graviton-graviton'' contributions ($U_{\rm GG}^{\rm in}$, respectively $U_{\rm GG}^{\rm out}$)
are positive. 
As expected, the total gravitational energy is found to be negative for all solutions, regardless of the compactness
of the source, the width of the gaussian, or the polytropic index. 
The values for each of these components for the cases discussed in Sections~\ref{S:numerical} and~\ref{S:gaussian}
can be found in Tables~\ref{t:grav_energy3} and~\ref{t:grav_energy4}, respectively. 
\begin{table}[t]
\centering
\begin{tabular}{||c|c|c||c|c|c||c||}
\hline\hline
$X$ & $\bar\gamma$ & $n$ & $U_{\rm BG}/M$ & $U_{\rm GG}^{\rm in}/M$ & $U_{\rm GG}^{\rm out}/M$  & $U_{\rm G}/M$ \\ \hline
\hline
$0.65$ & $1$ & $5/3$ & $-6.9$ & $3.6$ & $0.21$ & $-3.1$ \\ 
\hline
$0.55$ & $1$ & $5/3$ & $-5.8$ & $2.8$ & $0.15$ & $-2.8$ \\ 
\hline
$0.45$ & $1$ & $5/3$ & $-4.7$ & $2.0$ & $0.10$ & $-2.5$ \\ 
\hline
$0.10$ & $0.2$ & $5/3$ & $-1.7\times 10^{-2}$ & $4.0\times 10^{-3}$ & $4.0\times 10^{-3}$ & $-7.6\times 10^{-3}$ \\ 
\hline
$0.08$ & $0.2$ & $5/3$ & $-1.3\times 10^{-2}$ & $2.8\times 10^{-3}$ & $3.2\times 10^{-3}$ & $-7.3\times 10^{-3}$ \\ 
\hline
$0.06$ & $0.2$ & $5/3$ & $-1.0\times 10^{-2}$ & $1.8\times 10^{-3}$ & $1.8\times 10^{-3}$ & $-6.4\times 10^{-3}$ \\ 
\hline 
$0.65$ & $0.5$ & $4/3$ & $-6.6$ & $3.8$ & $0.21$ & $-2.5$ \\ 
\hline
$0.55$ & $0.5$ & $4/3$ & $-5.9$ & $3.3$ & $0.15$ & $-2.4$ \\ 
\hline
$0.45$ & $0.5$ & $4/3$ & $-5.2$ & $2.8$ & $0.10$ & $-2.3$ \\ 
\hline
\hline
\end{tabular}
\caption{Gravitational potential energy for the combinations of compactness and polytropic parameters
analysed in Section~\ref{S:numerical}.}
\label{t:grav_energy3}
\end{table}
\begin{table}
\centering
\begin{tabular}{||c|c|c||c|c|c||c||}
\hline\hline
$b$ & $X$ & $n$ & $U_{\rm BG}/M$ & $U_{\rm GG}^{\rm in}/M$ & $U_{\rm GG}^{\rm out}/M$  & $U_{\rm G}/M$ \\ \hline
\hline
0.5 & 0.01 & 3/2 & $-2.1 \times 10^{-4}$ & $4.9 \times 10^{-5}$ & $5.0 \times 10^{-5}$ & $-1.1 \times 10^{-4}$ \\ 
\hline
0.5 & 0.1 & 3/2 & $-2.0 \times 10^{-2}$ & $5.2\times 10^{-3}$ & $5.0\times 10^{-3}$ & $-9.7\times 10^{-3}$ \\ 
\hline
0.5 & 0.7 & 5/3 & -1.5 & 0.52 & 0.25 & -0.74 \\ 
\hline
1.0 & 0.01 & 3/2 & $-1.2 \times 10^{-4}$ & $1.3 \times 10^{-5}$ & $5.0 \times 10^{-5}$ & $-5.5 \times 10^{-5}$ \\ 
\hline
1.0 & 0.1 & 3/2 &$-1.1\times 10^{-2}$ & $1.4\times 10^{-3}$ & $5.0\times 10^{-3}$ & $-4.7\times 10^{-3}$ \\ 
\hline
1.0 & 0.7 & 5/3 & -0.52 & 0.07 & 0.25 & -0.20 \\ 
\hline
\hline
\end{tabular}
\caption{Gravitational potential energy for the combinations of Gaussian width, compactness, and polytropic index
analysed in Section~\ref{S:gaussian}.}
\label{t:grav_energy4}
\end{table}
\end{document}